\documentclass[preprintnumbers,superscriptaddress,nofootinbib,aps,prd,floatfix,notitlepage]{revtex4-1}

\usepackage{hyperref}
\usepackage{amsmath,slashed}
\usepackage{graphicx}
\usepackage{caption}
\usepackage{subcaption}
\usepackage{relsize}
\usepackage{mathtools}

\usepackage{empheq}
\usepackage{amssymb}

\newcommand{\be}{\begin{eqnarray*}}
\newcommand{\ee}{\end{eqnarray*}}

\newcommand{\bee}{\begin{eqnarray}}
\newcommand{\eee}{\end{eqnarray}}
\newcommand{\beeq}{\begin{equation}}
\newcommand{\eeeq}{\end{equation}}

\newcommand{\mksmall}[1]{\textnormal{\smaller$#1$}}

\newcommand{\obs}[1]{\mathrm{O}_{#1}}

\newcommand{\one}[2]{\mathds{1}^{#1}_{#2}}

\newcommand{\nnnlo}{\textnormal{\smaller N3LO}}
\newcommand{\nnlo}{\textnormal{\smaller NNLO}}
\newcommand{\nlo}{\textnormal{\smaller NLO}}
\newcommand{\lo}{\textnormal{\smaller LO}}
\newcommand{\qcd}{\textnormal{\smaller QCD}}
\newcommand{\pdf}{\textnormal{\smaller PDF}}
\newcommand{\lhc}{\textnormal{\smaller LHC}}
\newcommand{\dglap}{\textnormal{\smaller DGLAP}}

\newcommand{\xf}[3]{#2 f_{#1}\left(#2, #3\right)}

\newcommand{\pskernel}[3]{K_{#1\rightarrow #2}\left(#3\right)}

\newcommand{\lodglapkernel}[3]{P_{#1\leftarrow #2}^{(0)}\left(#3\right)}
\newcommand{\nlodglapkernel}[3]{P_{#1\leftarrow #2}^{(1)}\left(#3\right)}

\newcommand{\lopskernel}[3]{K_{#1\rightarrow #2}^{(0)}\left(#3\right)}
\newcommand{\nlopskernel}[3]{K_{#1\rightarrow #2}^{(1)}\left(#3\right)}

\newcommand{\nnlopps}{\textnormal{{\smaller NNLO}+{\smaller PS}}}
\newcommand{\nnnlopps}{\textnormal{{\smaller N3LO}+{\smaller PS}}}

\newcommand{\nklopps}{\textnormal{{\smaller N$^k$LO}+{\smaller PS}}}

\let\ss\undefined
\let\i\undefined

\newcommand{\ss}[4]{ d\sigma_{#1}^{(#2) \left[\textnormal{\smaller$#3$}\right]}(#4)}

\newcommand{\s}[3]{ d\sigma_{#1}^{(#2)}(#3)}

\newcommand{\i}[1]{ \int \mathrm{d}\Phi_{#1}}

\newcommand{\unlops}{\textnormal{\smaller UNLOPS}}

\newcommand{\nnnlops}{\textnormal{\textsc{Tomte}}}

\DeclareGraphicsExtensions{.pdf,.png,.jpg}

\usepackage{tikz}
\usetikzlibrary{tikzmark}
\usetikzlibrary{calc} 
\usetikzlibrary{decorations.pathmorphing}
\usetikzlibrary{shapes.symbols}
\usetikzlibrary{backgrounds}

\usepackage{tikz-feynman}
\tikzfeynmanset{compat=1.0.0}

 \tikzset{oplus/.style={path picture={%
 \begin{pgfonlayer}{background} 
 \draw[red,opacity=.5]
  (path picture bounding box.south east) -- (path picture bounding box.north west) 
  (path picture bounding box.south west) -- (path picture bounding box.north east);
\end{pgfonlayer}}}
}

\tikzstyle{description} = [rectangle, minimum width=1cm, minimum height=0.5cm, draw=black, fill=green!05, inner sep=2pt, inner ysep=2pt, inner xsep=2pt]
\tikzstyle{nobox} = [rectangle, minimum width=0cm, minimum height=0.0cm, text centered, draw=black!00, fill=black!00, inner sep=1pt, inner ysep=1pt]
\tikzstyle{box} = [rectangle, minimum width=0cm, minimum height=0.0cm, text centered, draw=black, fill=black!00, inner sep=1pt, inner ysep=1pt]
\tikzstyle{circ} = [circle, minimum width=0cm, minimum height=0.0cm, text centered, draw=black, fill=black!00, inner sep=1pt, inner ysep=1pt]
\tikzstyle{arrow} = [thick,->,>=stealth]
\tikzstyle{fixedbox15} = [rectangle, minimum width=1.5cm, minimum height=1.5cm, text centered, draw=black, fill=black!00, inner sep=1pt, inner ysep=1pt]
\tikzstyle{fixedbox11} = [rectangle, minimum width=1.1cm, minimum height=1.1cm, text centered, draw=black, fill=black!00, inner sep=1pt, inner ysep=1pt]
\tikzstyle{fixedbox10} = [rectangle, minimum width=1.0cm, minimum height=1.0cm, text centered, draw=black, fill=black!00, inner sep=1pt, inner ysep=1pt]
\tikzstyle{fixedbox09} = [rectangle, minimum width=0.9cm, minimum height=0.9cm, text centered, draw=black, fill=black!00, inner sep=1pt, inner ysep=1pt]
\tikzstyle{fixedbox08} = [rectangle, minimum width=0.8cm, minimum height=0.8cm, text centered, draw=black, fill=black!00, inner sep=1pt, inner ysep=1pt]
\tikzstyle{fixedbox07} = [rectangle, minimum width=0.7cm, minimum height=0.7cm, text centered, draw=black, fill=black!00, inner sep=1pt, inner ysep=1pt]
\tikzstyle{fixedbox06} = [rectangle, minimum width=0.6cm, minimum height=0.6cm, text centered, draw=black, fill=black!00, inner sep=1pt, inner ysep=1pt]
\tikzstyle{fixedbox05} = [rectangle, minimum width=0.5cm, minimum height=0.5cm, text centered, draw=black, fill=black!00, inner sep=1pt, inner ysep=1pt]

\usepackage{tcolorbox}
\tcbuselibrary{theorems}

\tcbsetforeverylayer{arc=0pt,outer arc=0pt,top=0pt,bottom=0pt,left=0pt,right=0pt,colback=black!10,colframe=black, boxrule=0.1pt, boxsep=0.1pt}
\tcbuselibrary{skins}
\tcbsetforeverylayer{skin=enhanced}

\usepackage{amsfonts}

\usepackage{color}

\usepackage{cleveref}
\usepackage{dsfont}

\definecolor{mypurple}{rgb}{0.8,0.1,0.7}

\definecolor{myolive}{rgb}{0.2,0.45,0.25}

\definecolor{mybrown}{rgb}{0.54,0.27,0.07}

\definecolor{mymidgrey}{rgb}{0.5,0.5,0.5}

\definecolor{mywhite}{rgb}{1.0,1.0,1.0}



\allowdisplaybreaks

\makeatletter
\def\upintkern@{\mkern-7mu\mathchoice{\mkern-3.5mu}{}{}{}}
\def\upintdots@{\mathchoice{\mkern-4mu\@cdots\mkern-4mu}%
 {{\cdotp}\mkern1.5mu{\cdotp}\mkern1.5mu{\cdotp}}%
 {{\cdotp}\mkern1mu{\cdotp}\mkern1mu{\cdotp}}%
 {{\cdotp}\mkern1mu{\cdotp}\mkern1mu{\cdotp}}}

\newcommand{\UpMultiIntegral}[1]{%
  \edef\ints@c{\noexpand\upintop
    \ifnum#1=\z@\noexpand\upintdots@\else\noexpand\upintkern@\fi
    \ifnum#1>\tw@\noexpand\upintop\noexpand\upintkern@\fi
    \ifnum#1>\thr@@\noexpand\upintop\noexpand\upintkern@\fi
    \noexpand\upintop
    \noexpand\ilimits@
  }%
  \futurelet\@let@token\ints@a
}
\makeatother

\DeclareFontFamily{OMX}{mdbch}{}
\DeclareFontShape{OMX}{mdbch}{m}{n}{ <->s * [0.8]  mdbchr7v }{}
\DeclareFontShape{OMX}{mdbch}{b}{n}{ <->s * [0.8]  mdbchb7v }{}
\DeclareFontShape{OMX}{mdbch}{bx}{n}{<->ssub * mdbch/b/n}{}

\DeclareSymbolFont{uplargesymbols}{OMX}{mdbch}{m}{n}
\SetSymbolFont{uplargesymbols}{bold}{OMX}{mdbch}{b}{n}
\DeclareMathSymbol{\upintop}{\mathop}{uplargesymbols}{82}
\DeclareMathSymbol{\upointop}{\mathop}{uplargesymbols}{"48}

\DeclareFontEncoding{MDB}{}{}
\DeclareFontFamily{MDB}{mdbch}{}
\DeclareFontShape{MDB}{mdbch}{m}{n}{ <->s * [0.8]  mdbchrmb }{}
\DeclareFontShape{MDB}{mdbch}{b}{n}{ <->s * [0.8]  mdbchbmb }{}
\DeclareFontShape{MDB}{mdbch}{bx}{n}{<->ssub * mdbch/b/n}{}
\DeclareFontSubstitution{MDB}{cmr}{m}{n}
\DeclareSymbolFont{mathdesignB}{MDB}{mdbch}{m}{n}%
\SetSymbolFont{mathdesignB}{bold}{MDB}{mdbch}{b}{n}%
\DeclareMathSymbol{\upintclockwise}{\mathop}{mathdesignB}{128}
\DeclareMathSymbol{\upointclockwise}{\mathop}{mathdesignB}{130}
\DeclareMathSymbol{\upointctrclockwise}{\mathop}{mathdesignB}{132}
\DeclareMathSymbol{\upoiint}{\mathop}{mathdesignB}{134}
\DeclareMathSymbol{\upoiiint}{\mathop}{mathdesignB}{136}

\makeatletter
\newcommand{\upint}{\DOTSI\upintop\ilimits@}
\newcommand{\upoint}{\DOTSI\upointop\ilimits@}

\renewcommand{\int}{\mathop{\mathlarger{\upint}}}

\begin{document}

\title{Combining N3LO QCD calculations and parton showers for hadronic collision events}

\begin{abstract}
%\abstract{
Detailed and precise background predictions are the backbone of large parts of high-energy collider phenomenology. This requires to embed precision \qcd{} calculations into detailed event generators, to produce comprehensive software simulations. Only continued progress in this direction will allow us to exploit the full potential of measurements at the Large Hadron Collider, or at a future Electron-Ion Collider. This work presents a method to combine third-order \qcd{} calculations for hadronic scattering processes with Monte-Carlo event generators, thus enabling a new generation of precision predictions.
%}
\end{abstract}

%\author[a]{Stefan Prestel}
%\affiliation[a]{Department of Astronomy and Theoretical Physics,\\Lund University, S-223 62 Lund, Sweden}

%\emailAdd{stefan.prestel@thep.lu.se}

\author{Valerio Bertone}
\affiliation{Irfu, CEA, Université Paris-Saclay, 91191, Gif-sur-Yvette, France}

\author{Stefan Prestel}
\affiliation{Department of Astronomy and Theoretical Physics,\\Lund University, S-223 62 Lund, Sweden}

%comment out for jhep

\preprint{LU-TP-22-06}
\vspace*{4ex}

%\maketitle

% avoid new page after title page
\maketitle

\section{Introduction}

Measurements at particle colliders aim to provide insights into the fundamental building blocks of physics by juxtaposing experimentally recorded scattering final states with detailed theory simulations. Any deviation from the expectation (based on the Standard Model of particle physics) hints at new research directions.

This ``indirect search" strategy relies on sophisticated event generators, which should furnish an accurate model of the scattering dynamics~\cite{Buckley:2011ms}. On top of this, precision predictions have become ever more important, as detailed error budgets are mandatory for reliable comparisons to experimental data. Such predictions are crucial to the \lhc{} phenomenology programme -- where new-physics signals have to be lifted from immense \qcd{} backgrounds -- as well as future Electron-Ion colliders, where novel \qcd{} phenomena have to be confronted with higher-order \qcd{} calculations within the collinear factorization approximation. These simulations rely on combining high-precision fixed-order (\qcd{}) calculations with parton evolution via all-order parton showering, yielding higher-order event generators.

\nlo{} event generators~\cite{Frixione:2002ik,*Nason:2004rx,*Frixione:2007vw} have become the staple of \lhc{} phenomenology, while \nnlo{} event generators have, albeit still requiring cutting-edge research, become more commonplace~\cite{Lavesson:2008ah,*Hoeche:2014aia,*Hoche:2014dla,*Hamilton:2013fea,*Karlberg:2014qua,*Hamilton:2015nsa,*Alioli:2015toa,*Astill:2018ivh,*Hoche:2018gti,*Re:2018vac,*Monni:2019whf,*Monni:2020nks,*Lombardi:2020wju,*Hu:2021rkt,*Alioli:2020qrd,*Mazzitelli:2020jio}. These generators typically require dedicated implementations of higher-order calculations. Thus, methods to enable higher-order event generators should ideally be known when leaps in precision are achieved in fixed-order calculations. 

Recent years have seen impressive progress in producing \nnnlo{} fixed-order \qcd{} predictions, both at the inclusive~\cite{Anastasiou:2015vya,*Duhr:2019kwi,*Duhr:2020seh,*Chen:2019lzz,*Duhr:2020sdp} and the fully differential level~\cite{Dulat:2017prg,*Currie:2018fgr,*Dreyer:2018qbw,*Cieri:2018oms,*Mondini:2019gid,*Chen:2021isd,Billis:2021ecs,Camarda:2021ict,Re:2021con}. In some cases, these fully-differential results have even been used to supplement resummed predictions for important observables~\cite{Banfi:2015pju,Billis:2021ecs,Camarda:2021ict,Re:2021con}. This note offers a method to produce \nnnlo{} event generators for processes with incoming hadrons, expanding on the proof-of-principle work~\cite{Prestel:2021vww} concerned with leptonic collisions. The method employs an intuitive \emph{``subtract what you add"} scheme to disentangle fixed-order corrections and parton-shower contributions, up to third order in \qcd{}. The feasibility of an implementation of the method is tested using the controlled environment of a ``toy \nnnlo{} calculation"\footnote{Differential \nnnlo{} fixed-order calculations that produce finite-weight \emph{events} have yet to emerge.}. Overall, these tests provide abstract arguments for the validity and accuracy of the \nnlopps{} method with a numerical verification. 

\section{N3LO matching}

High-precision event generators typically rely on \emph{matching} or \emph{merging} schemes, which first define the desired precision of the simulation (potentially for specific observables), and then provide an algorithmic realization by combining (possibly a set of) precise fixed-order calculations with subsequent parton showering, while ensuring that no terms are double counted. This allows one to improve both on the accuracy of the simulation -- by containing a better approximation of (real-emission) multi-particle states compared to showering -- and the precision of the simulation by containing exact virtual corrections, thus (hopefully) yielding better control over renormalization scale uncertainties. A successful matching scheme will ensure that all terms up to the desired order are correctly described at fixed order, while all higher-order terms reproduce the parton-shower result. The latter is ambiguous. This note will follow the reasoning and shower accuracy definition of~\cite{Prestel:2021vww}, which is based on obtaining a self-consistent calculation without approximating the shower result by an (observable-dependent) log-counting. That being said, we expect the parton-shower (no-emission probabilities) to regularize cross sections of ordered, single-unresolved emissions in the soft and collinear limits. 

\begin{figure}[tbp]
\centering
  \includegraphics[width=0.6\textwidth]{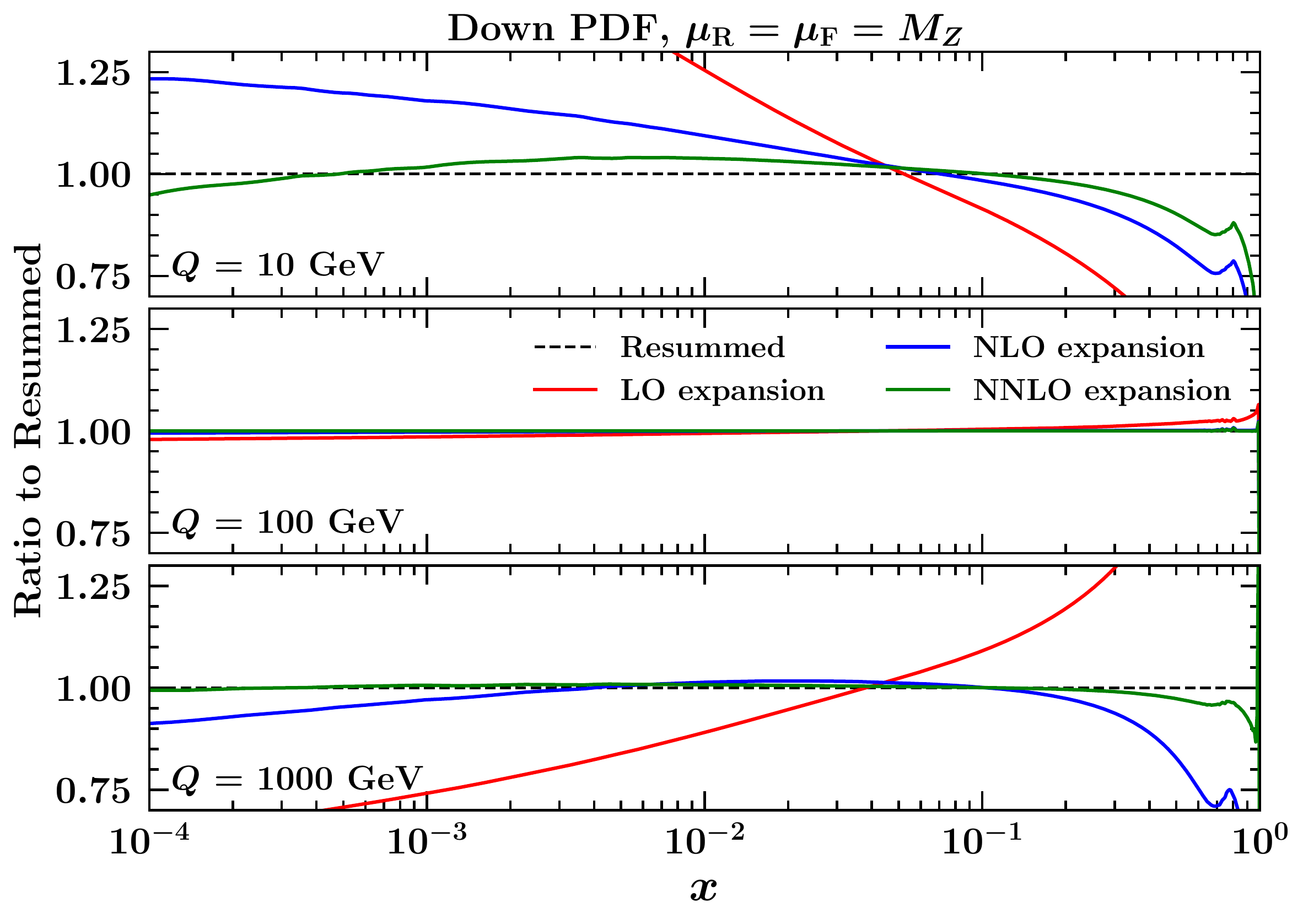}{}
  \caption{\label{fig:pdfexp} Comparison of resummed vs. expanded
    solution of the \dglap{} evolution equations. The down-quark \pdf{} as a
    function of $x$ at three different values of the scale $Q=$ 10~GeV
    (upper plot), 100~GeV (central plot), and 1000~GeV (lower plot) is
    computed in terms of the \pdf{}s at scale the $\mu_{\rm F}$ and
    $\alpha_s(\mu_R)$, with $\mu_{\rm R}=\mu_{\rm F}=M_Z$, using the
    resummed solution (dashed black lines) and its expansion to three
    different orders (see
    Eqs.~(\ref{eq:pdfevolutionexpansionLOandNLO})-(\ref{eq:pdfevolutionexpansionNNLO})):
    \lo{} (red curves), \nlo{} (blue curves), \nnlo{} (green curves). The
    curves are presented as ratios to the resummed solution.}
\end{figure}

Typically, the fixed-order and shower regions are regarded as basically complementary. However, very high-precision fixed-order simulations may in fact already provide an approximation of all-order perturbative effects that is -- although only an effective description -- superior to truly all-order, yet approximate, description of parton showers. One such all-order effect is the evolution of parton distribution functions in initial state showers. Common showering algorithms based on backward evolution e.g.\ fail to recover \dglap{} evolution for long evolution without emission (see contribution titled ``Self-consistency of backwards evolved initial-state parton showers" in \cite{Amoroso:2020lgh}\footnote{Methods to overcome such issues have been discussed~\cite{Gottschalk:1986bk,Nagy:2020gjv}, but require a radical redesign of showering algorithms}). However, already using the (correct) second-order expansion of \pdf{} evolution yields a satisfactory approximation of the all-order result, as shown in Figure~\ref{fig:pdfexp}. Thus, replacing the coefficients of the expansion of the parton shower up to $\mathcal{O}(\alpha_s^2)$ with the correct values will help ameliorate problems with \pdf{} evolution. Only \nnnlopps{} methods \emph{demand} that the $\mathcal{O}(\alpha_s^2)$ coefficients of the shower be replaced, making hadron-collider extensions of the \nnnlops{} scheme desirable.

The goal of \nnnlopps{} methods is to provide a unified calculation that retains the relevant fixed-order precision for up to three additional partons throughout all of phase space, while supplementing the fixed-order expansion with all-order effects provided by the parton shower. The accuracy and internal consistency of the parton-shower resummation must not be affected by the matching prescription. Observables that do not require the support of any parton should be described at \nnnlo{} precision, observables demanding the presence of at least one parton should be described at \nnlo{} precision, observables mandating two additional final-state partons with \nlo{} precision, and observables depending on at least three additional partons 
should be predicted with \lo{} accuracy. Whenever any observable is ``exclusive", i.e.\ relies on a fixed number of final-state jets, all-order effects beyond $\mathcal{O}(\alpha_s^3)$ should be calculated using the parton shower resummation.

%\begin{itemize}
%\item general intro, why matching?
%\item matching allows to control/improve the uncertainties
%\item matching makes the development of improved event generators more democratic, since it ideally does not require generator author knowledge
%\item if matching is of high enough order, matching can already provide an improved ``effective" approximation of all-order (PDF) evolution effects.
%\end{itemize}

\subsection{Basic concepts}
\label{sec:basics}

This section aims to introduce the \nnnlops{} method, and illustrate its derivation and most important principles. The original \nnnlops{} publication~\cite{Prestel:2021vww} developed the method in full generality, providing detailed definitions, intermediate formulae, and somewhat length end results. To complement this discussion, this note will provide a diagrammatic explanation of the method. For this, let us first introduce diagrams that will feature heavily later. The symbol
\begin{equation*}
\textnormal{} 
\quad
\vcenter{\hbox{
\begin{tikzpicture}[remember picture]
\node (def) [nobox,inner sep=2pt, inner ysep=2pt,scale=0.6] {
\begin{tikzpicture}%
  \begin{feynman} [every blob={/tikz/fill=gray!10,/tikz/inner sep=2pt}]%
    \vertex [xshift=0.0cm] (a) {\smaller };%
    \vertex [right=of a, blob] (b) {\smaller };%
    %\vertex [below=of b, yshift=0.5cm] (glu1) {\smaller \phantom{glu1}};
    \vertex [right=of b] (c) {\smaller };%
    %\vertex [above=of b,yshift=-0.6cm, align=center] (title) {\smaller Hard process};
    \diagram* { (a) -- (b), (b) -- (c); };%
  \end{feynman}%
\end{tikzpicture}%
};
\end{tikzpicture}%
}}
\quad
\textnormal{} 
\end{equation*}
indicates an arbitrary lowest-multiplicity process. Zeroth-order (tree-level) corrections with $n$ additional resolved partons to this process will be illustrated by
\begin{equation*}
\textnormal{} 
\quad
\vcenter{\hbox{
\begin{tikzpicture}[remember picture]
\node (def) [nobox,inner sep=2pt, inner ysep=2pt,scale=0.6] {
\begin{tikzpicture}
  \begin{feynman} [every blob={/tikz/fill=gray!10,/tikz/inner sep=2pt}]
    \vertex [xshift=0.0cm] (a) {\smaller };
    \vertex [right=of a, blob] (b) {\smaller };
    \vertex [circ,right=of b, xshift=0.0cm] (v1) {\smaller };
    \vertex [below=of v1, yshift=0.5cm] (glu1) {\smaller 1};
    \vertex [circ,right=of v1, xshift=0.0cm] (v2) {\smaller };
    \vertex [below=of v2, yshift=0.5cm] (glu2) {\smaller $n$};
    \vertex [right=of v2] (c) {\smaller };
    \diagram* {
      (a) -- (b),
      (b) -- (v1),
      (v1) -- (v2),
      (v2) -- (c),
      (v1) -- [gluon] (glu1),
      (v2) -- [gluon] (glu2),
      (glu1) -- [ghost] (glu2)
    };
  \end{feynman}
\end{tikzpicture}
};
\end{tikzpicture}%
}}
\quad
\textnormal{,} 
\end{equation*}
while exclusive first-order (\nlo{}) corrections with $n$ additional partons (comprising one-loop corrections and unresolved real-emission contributions) will be given by
\begin{equation*}
\textnormal{} 
\quad
\vcenter{\hbox{
\begin{tikzpicture}[remember picture]
\node (def) [nobox,inner sep=2pt, inner ysep=2pt,scale=0.6] {
\begin{tikzpicture}
  \begin{feynman} [every blob={/tikz/fill=gray!10,/tikz/inner sep=2pt}]
    \vertex [xshift=-0.5cm] (a) {\smaller };
    \vertex [circ,right=of a] (v0) {\smaller };
    \vertex [right=of v0, blob] (b) {\smaller };
    \vertex [circ,right=of b, xshift=0.0cm] (v1) {\smaller };
    \vertex [below=of v1, yshift=0.5cm] (glu1) {\smaller 1};
    \vertex [circ,right=of v1, xshift=0.0cm] (v2) {\smaller };
    \vertex [below=of v2, yshift=0.5cm] (glu2) {\smaller n};
    \vertex [right=of v2] (c) {\smaller };
    %\vertex [above=of v1,yshift=-0.6cm] (title) {\smaller Exclusive first-order +$n$-jet corrections:};
    \diagram* {
      (a) -- (v0),
      (v0) -- (b),
      (b) -- (v1),
      (v1) -- [gluon] (glu1),
      (v1) -- (v2),
      (v2) -- [gluon] (glu2),
      (glu1) -- [ghost] (glu2),
      (b) -- [gluon, quarter right] (v0),
      (v2) -- (c)
    };
  \end{feynman}
\end{tikzpicture}
};
\end{tikzpicture}%
}}
\quad
\textnormal{.} 
\end{equation*}
Similarly, exclusive second-order (\nnlo{}) +$n$-jet corrections will be depicted by
\begin{equation*}
\textnormal{} 
\quad
\vcenter{\hbox{
\begin{tikzpicture}[remember picture]
\node (def) [nobox,inner sep=2pt, inner ysep=2pt,scale=0.6] {
\begin{tikzpicture}
  \begin{feynman} [every blob={/tikz/fill=gray!10,/tikz/inner sep=2pt}]
    \vertex [xshift=-0.5cm] (a) {\smaller };
    \vertex [circ,right=of a] (v0) {\smaller };
    \vertex [right=of v0, blob] (b) {\smaller };
    \vertex [circ,right=of b, xshift=0.0cm] (v1) {\smaller };
    \vertex [below=of v1, yshift=0.5cm] (glu1) {\smaller 1};
    \vertex [circ,right=of v1, xshift=0.0cm] (v2) {\smaller };
    \vertex [circ,right=of v2, xshift=-0.8cm] (v3) {\smaller };
    \vertex [below=of v3, yshift=0.5cm] (glu3) {\smaller $n$};
    \vertex [right=of v3] (c) {\smaller };
    %\vertex [above=of v1,yshift=-0.2cm,align=center] (title) {\smaller Exclusive second-order +$n$-jet corrections:};
    \diagram* {
      (a) -- (v0),
      (v0) -- (b),
      (b) -- (v1),
      (v1) -- (v2),
      (v2) -- (v3),
      (v1) -- [gluon] (glu1),
      (b) -- [gluon, quarter left] (v2),
      (b) -- [gluon, quarter right] (v0),
      (v3) -- [gluon] (glu3),
      (v3) -- (c),
      (glu1) -- [ghost] (glu3)
    };
  \end{feynman}
\end{tikzpicture}
};
\end{tikzpicture}%
}}
\quad
\textnormal{,} 
\end{equation*}
and exclusive third-order (\nnnlo{}) +$n$-jet corrections given by
\begin{equation*}
\textnormal{} 
\quad
\vcenter{\hbox{
\begin{tikzpicture}[remember picture]
\node (def) [nobox,inner sep=2pt, inner ysep=2pt,scale=0.6] {
\begin{tikzpicture}
  \begin{feynman} [every blob={/tikz/fill=gray!10,/tikz/inner sep=2pt}]
    \vertex [xshift=-0.5cm] (a) {\smaller };
    \vertex [circ,right=of a] (v0) {\smaller };
    \vertex [right=of v0, blob] (b) {\smaller };
    \vertex [right=of b,xshift=4.0cm] (c) {\smaller };
    \vertex [circ,right=of b, xshift=0.0cm] (v1) {\smaller };
    \vertex [circ,right=of v1, xshift=-0.7cm] (v11) {\smaller };
    \vertex [below=of v11, yshift=0.5cm] (glu11) {\smaller 1};
    \vertex [circ,right=of v11, xshift=0.0cm] (v2) {\smaller };
    \vertex [below=of v2, yshift=0.5cm] (glu2) {\smaller };
    \vertex [circ,right=of v2, xshift=-0.8cm] (v3) {\smaller };
    \vertex [below=of v3, yshift=0.5cm] (glu3) {\smaller $n$};
    \vertex [right=of v3] (c) {\smaller };
%    \vertex [above=of v1,yshift=-0.2cm,align=center] (title) {\smaller Exclusive third-order +$n$-jet corrections:};
    \diagram* {
      (a) -- (v0),
      (v0) -- (b),
      (b) -- (v1),
      (v1) -- (v11),
      (v11) -- (v2),
      (v2) -- (v3),
      (v3) -- (c),
      (b) -- [gluon, quarter right] (v1),
      (b) -- [gluon, quarter left] (v2),
      (b) -- [gluon, quarter right] (v0),
      (v11) -- [gluon] (glu11),
      (v3) -- [gluon] (glu3),
      (glu11) -- [ghost] (glu3),
    };
  \end{feynman}
\end{tikzpicture}
};
\end{tikzpicture}%
}}
\quad
\textnormal{.} 
\end{equation*}
Symbols describing the action of the parton shower include the shower emission vertex
\begin{equation*}
\textnormal{} 
\quad
\vcenter{\hbox{
\begin{tikzpicture}[remember picture]
\node (def) [nobox,inner sep=2pt, inner ysep=2pt,scale=0.7] {
\begin{tikzpicture}
  \begin{feynman} [every blob={/tikz/fill=gray!10,/tikz/inner sep=2pt}]
    \vertex [xshift=0.0cm] (a) {\smaller };
    \vertex [dot, right=of a] (v1) {\smaller };
    %\vertex [below=of v1, yshift=0.5cm, align=center] (glu1) {\smaller soft/coll.\ emission at $t$};
    \vertex [below=of v1, yshift=0.5cm, align=center] (glu1) {\smaller $t$};
    \vertex [right=of v1] (c) {\smaller };
    %\vertex [above=of v1,yshift=-0.6cm, align=center] (title) {\smaller Shower vertex at $t$ };
    \diagram* {
      (a) -- (v1),
      (v1) -- (c),
      (v1) -- [scalar] (glu1)
    };
  \end{feynman}
\end{tikzpicture}
};
\end{tikzpicture}%
}}
\quad
\textnormal{.} 
\end{equation*}
This vertex encapsulates the (higher-order) effect of dynamical renormalization and factorization scale choices for shower emissions at evolution scale $t$. The probability of not emitting between two shower emissions at scales $t_1$ and $t_2$ is symbolized by multiple lines,
\begin{equation*}
\textnormal{} 
\quad
\vcenter{\hbox{
\begin{tikzpicture}[remember picture]
\node (def) [nobox,inner sep=2pt, inner ysep=2pt,scale=0.7] {
\begin{tikzpicture}
  \begin{feynman} [every blob={/tikz/fill=gray!10,/tikz/inner sep=2pt}]
    \vertex [] (a) {\smaller };
    \vertex [circ, right=of a, xshift=0.0cm] (v1) {\smaller };
    \vertex [below=of v1, yshift=0.5cm, align=center] (glu1) {\smaller $t_1$};
    \vertex [circ, right=of v1, xshift=0.0cm] (v2) {\smaller };
    \vertex [below=of v2, yshift=0.5cm, align=center] (glu2) {\smaller $t_2$};
    \vertex [right=of a, yshift=0.1cm] (v1u)  {\smaller };
    \vertex [right=of a, yshift=-0.1cm] (v1d) {\smaller };
    \vertex [right=of v1u] (v2u) {\smaller };
    \vertex [right=of v1d] (v2d) {\smaller };
    \vertex [right=of v2] (c) {\smaller };
    %\vertex [above=of v1,yshift=-0.6cm,xshift=0.8cm, align=center] (title) {\smaller No emission between $t_1$ and $t_2$ };
    \diagram* {
      (a) -- (v1),
      (v1) -- (v2),
      (v2) -- (c),
      (v1u) -- (v2u),
      (v1d) -- (v2d),
      (v1) -- [scalar] (glu1),
      (v2) -- [scalar] (glu2)
    };
  \end{feynman}
\end{tikzpicture}
};
\end{tikzpicture}%
}}
\quad
\textnormal{.} 
\end{equation*}
Finally, it will be necessary to depict contributions obtained by integrating over the degrees of freedom of one or several final-state particles:
\begin{equation*}
\textnormal{} 
\quad
\vcenter{\hbox{
\begin{tikzpicture}[remember picture]
\node (def) [nobox,inner sep=2pt, inner ysep=2pt,scale=0.7] {
\begin{tikzpicture}
  \begin{feynman} [every blob={/tikz/fill=gray!10,/tikz/inner sep=2pt}]
    \vertex [xshift=0.0cm] (a) {\smaller };
    \vertex [circ,right=of a] (v1) {\smaller };
    \vertex [crossed dot, below=of v1, yshift=0.5cm] (glu1) {\smaller \phantom{glu1}};
    \vertex [right=of v1] (c) {\smaller };
    %\vertex [above=of v1,yshift=-0.6cm, align=center] (title) {\smaller Gluon emission is integrated out};
    \diagram* {
      (a) -- (v1),
      (v1) -- (c),
      (v1) -- [gluon] (glu1)
    };
  \end{feynman}
\end{tikzpicture}
};
\end{tikzpicture}
}}
\quad
\textnormal{.} 
\end{equation*}
This integration is performed by sampling the full multi-parton phase space, and then projecting the configurations onto the desired multiplicity by choosing an underlying configuration, as first discussed in \cite{Rubin:2010xp,Lonnblad:2012ng}.

When constructing a third-order matched calculation, it will be useful to move from the description of high multiplicity states to that of low multiplicity states, as high-multiplicity states enter as real-emission corrections in lower-multiplicity calculations. Also, the handling of real-emission contributions in the parton shower is directly linked to the generation of no-emission probabilities by virtue of the unitarity of the showering process. In the following, we will (first) discuss the basic concepts of the \nnnlops{} method using the fixed-order contributions starting at $\mathcal{O}(\alpha_s^3)$.

A suitable treatment of the highest possible multiplicity contributions (containing three additional partons relative to the Born process) only requires the comparison of tree-level result $\s{n+3}{0}{\Phi_{n+3}}$ with its parton shower analogon
\begin{equation*}
\textnormal{} 
\quad
\vcenter{\hbox{
\begin{tikzpicture}[remember picture]
\node (lops3) [nobox,inner sep=2pt, inner ysep=2pt,scale=0.7] {
\begin{tikzpicture}
  \begin{feynman} [every blob={/tikz/fill=gray!10,/tikz/inner sep=2pt}]
    \vertex [xshift=-0.5cm] (a) {\smaller };
    \vertex [right=of a] (v0) {\smaller };
    \vertex [right=of v0, blob] (b) {\smaller };
    \vertex [] (bu) at ($(b)+(0.27,0.1)$) {\smaller } ;
    \vertex [] (bd) at ($(b)+(0.27,-0.1)$) {\smaller };
    \vertex [circ,right=of b, xshift=0.0cm] (v1) {\smaller };
    \vertex [below=of v1, yshift=0.5cm] (glu1) {\smaller \phantom{$t_1$}};
    \vertex [circ,right=of v1, xshift=0.0cm] (v2) {\smaller };
    \vertex [below=of v2, yshift=0.5cm] (glu2) {\smaller };
    \vertex [circ,right=of v2, xshift=0.0cm] (v3) {\smaller };
    \vertex [below=of v3, yshift=0.5cm] (glu3) {\smaller };
    \vertex [] (v1u) at ($(v1)+(0.0,0.1)$) {\smaller };
    \vertex [] (v1d) at ($(v1)+(-0.07,-0.1)$) {\smaller };
    \vertex [right=of v1u] (v2u) {\smaller };
    \vertex [right=of v1d] (v2d) {\smaller };
    \vertex [right=of v2u] (v3u) {\smaller };
    \vertex [right=of v2d] (v3d) {\smaller };
    \vertex [right=of v3] (c) {\smaller };
    %\vertex [above=of b,yshift=-0.6cm,xshift=-1.0cm,align=left] (title) {\smaller Reweighted tree-level\\ three-jet contributions};
    \diagram* {
      (a) -- (b),
      (b) -- (v1),
      (v1) -- (v2),
      (v2) -- (v3),
      (v3) -- (c),
      %(bu) -- (v1u),
      %(bd) -- (v1d),
      %(v1u) -- (v2u),
      %(v1d) -- (v2d),
      %(v2u) -- (v3u),
      %(v2d) -- (v3d),
      (v1) -- [gluon] (glu1),
      (v2) -- [gluon] (glu2),
      (v3) -- [gluon] (glu3),
      %(v1) -- [gluon, quarter right] (glu2) -- [gluon, quarter right] (v3),
    };
  \end{feynman}
\end{tikzpicture}
};
\end{tikzpicture}
}}
\quad
\textnormal{vs.} 
\quad
\vcenter{\hbox{
\begin{tikzpicture}[remember picture]
\node (lops3) [nobox,inner sep=2pt, inner ysep=2pt,scale=0.7] {
\begin{tikzpicture}
  \begin{feynman} [every blob={/tikz/fill=gray!10,/tikz/inner sep=2pt}]
    \vertex [xshift=-0.5cm] (a) {\smaller };
    \vertex [right=of a] (v0) {\smaller };
    \vertex [right=of v0, blob] (b) {\smaller };
    \vertex [] (bu) at ($(b)+(0.27,0.1)$) {\smaller } ;
    \vertex [] (bd) at ($(b)+(0.27,-0.1)$) {\smaller };
    \vertex [dot,right=of b, xshift=0.0cm] (v1) {\smaller };
    \vertex [below=of v1, yshift=0.5cm] (glu1) {\smaller $t_1$};
    \vertex [dot,right=of v1, xshift=0.0cm] (v2) {\smaller };
    \vertex [below=of v2, yshift=0.5cm] (glu2) {\smaller $t_2$};
    \vertex [dot,right=of v2, xshift=0.0cm] (v3) {\smaller };
    \vertex [below=of v3, yshift=0.5cm] (glu3) {\smaller $t_3$};
    \vertex [] (v1u) at ($(v1)+(0.0,0.1)$) {\smaller };
    \vertex [] (v1d) at ($(v1)+(-0.07,-0.1)$) {\smaller };
    \vertex [right=of v1u] (v2u) {\smaller };
    \vertex [right=of v1d] (v2d) {\smaller };
    \vertex [right=of v2u] (v3u) {\smaller };
    \vertex [right=of v2d] (v3d) {\smaller };
    \vertex [right=of v3] (c) {\smaller };
    %\vertex [above=of b,yshift=-0.6cm,xshift=-1.0cm,align=left] (title) {\smaller Reweighted tree-level\\ three-jet contributions};
    \diagram* {
      (a) -- (b),
      (b) -- (c),
      (bu) -- (v1u),
      (bd) -- (v1d),
      (v1u) -- (v2u),
      (v1d) -- (v2d),
      (v2u) -- (v3u),
      (v2d) -- (v3d),
      (v1) -- [scalar] (glu1),
      (v2) -- [scalar] (glu2),
      (v3) -- [scalar] (glu3),
      %(v1) -- [gluon, quarter right] (glu2) -- [gluon, quarter right] (v3),
    };
  \end{feynman}
\end{tikzpicture}
};
\end{tikzpicture}
}}
\end{equation*}
Thus, incorporating all shower higher-order corrections amounts to the replacement
\begin{equation*}
\textnormal{} 
\quad
\vcenter{\hbox{
\begin{tikzpicture}[remember picture]
\node (lops3) [nobox,inner sep=2pt, inner ysep=2pt,scale=0.7] {
\begin{tikzpicture}
  \begin{feynman} [every blob={/tikz/fill=gray!10,/tikz/inner sep=2pt}]
    \vertex [xshift=-0.5cm] (a) {\smaller };
    \vertex [right=of a] (v0) {\smaller };
    \vertex [right=of v0, blob] (b) {\smaller };
    \vertex [] (bu) at ($(b)+(0.27,0.1)$) {\smaller } ;
    \vertex [] (bd) at ($(b)+(0.27,-0.1)$) {\smaller };
    \vertex [circ,right=of b, xshift=0.0cm] (v1) {\smaller };
    \vertex [below=of v1, yshift=0.5cm] (glu1) {\smaller \phantom{$t_1$}};
    \vertex [circ,right=of v1, xshift=0.0cm] (v2) {\smaller };
    \vertex [below=of v2, yshift=0.5cm] (glu2) {\smaller };
    \vertex [circ,right=of v2, xshift=0.0cm] (v3) {\smaller };
    \vertex [below=of v3, yshift=0.5cm] (glu3) {\smaller };
    \vertex [] (v1u) at ($(v1)+(0.0,0.1)$) {\smaller };
    \vertex [] (v1d) at ($(v1)+(-0.07,-0.1)$) {\smaller };
    \vertex [right=of v1u] (v2u) {\smaller };
    \vertex [right=of v1d] (v2d) {\smaller };
    \vertex [right=of v2u] (v3u) {\smaller };
    \vertex [right=of v2d] (v3d) {\smaller };
    \vertex [right=of v3] (c) {\smaller };
    %\vertex [above=of b,yshift=-0.6cm,xshift=-1.0cm,align=left] (title) {\smaller Reweighted tree-level\\ three-jet contributions};
    \diagram* {
      (a) -- (b),
      (b) -- (v1),
      (v1) -- (v2),
      (v2) -- (v3),
      (v3) -- (c),
      %(bu) -- (v1u),
      %(bd) -- (v1d),
      %(v1u) -- (v2u),
      %(v1d) -- (v2d),
      %(v2u) -- (v3u),
      %(v2d) -- (v3d),
      (v1) -- [gluon] (glu1),
      (v2) -- [gluon] (glu2),
      (v3) -- [gluon] (glu3),
      %(v1) -- [gluon, quarter right] (glu2) -- [gluon, quarter right] (v3),
    };
  \end{feynman}
\end{tikzpicture}
};
\end{tikzpicture}
}}
\quad
\longrightarrow
\quad
\vcenter{\hbox{
\begin{tikzpicture}[remember picture]
\node (lops3) [nobox,inner sep=2pt, inner ysep=2pt,scale=0.7] {
\begin{tikzpicture}
  \begin{feynman} [every blob={/tikz/fill=gray!10,/tikz/inner sep=2pt}]
    \vertex [xshift=-0.5cm] (a) {\smaller };
    \vertex [right=of a] (v0) {\smaller };
    \vertex [right=of v0, blob] (b) {\smaller };
    \vertex [] (bu) at ($(b)+(0.27,0.1)$) {\smaller } ;
    \vertex [] (bd) at ($(b)+(0.27,-0.1)$) {\smaller };
    \vertex [dot,right=of b, xshift=0.0cm] (v1) {\smaller };
    \vertex [below=of v1, yshift=0.5cm] (glu1) {\smaller $t_1$};
    \vertex [dot,right=of v1, xshift=0.0cm] (v2) {\smaller };
    \vertex [below=of v2, yshift=0.5cm] (glu2) {\smaller $t_2$};
    \vertex [dot,right=of v2, xshift=0.0cm] (v3) {\smaller };
    \vertex [below=of v3, yshift=0.5cm] (glu3) {\smaller $t_3$};
    \vertex [] (v1u) at ($(v1)+(0.0,0.1)$) {\smaller };
    \vertex [] (v1d) at ($(v1)+(-0.07,-0.1)$) {\smaller };
    \vertex [right=of v1u] (v2u) {\smaller };
    \vertex [right=of v1d] (v2d) {\smaller };
    \vertex [right=of v2u] (v3u) {\smaller };
    \vertex [right=of v2d] (v3d) {\smaller };
    \vertex [right=of v3] (c) {\smaller };
    %\vertex [above=of b,yshift=-0.6cm,xshift=-1.0cm,align=left] (title) {\smaller Reweighted tree-level\\ three-jet contributions};
    \diagram* {
      (a) -- (b),
      (b) -- (c),
      (bu) -- (v1u),
      (bd) -- (v1d),
      (v1u) -- (v2u),
      (v1d) -- (v2d),
      (v2u) -- (v3u),
      (v2d) -- (v3d),
      (v1) -- [gluon] (glu1),
      (v2) -- [gluon] (glu2),
      (v3) -- [gluon] (glu3),
      %(v1) -- [gluon, quarter right] (glu2) -- [gluon, quarter right] (v3),
    };
  \end{feynman}
\end{tikzpicture}
};
\end{tikzpicture}
}}
\end{equation*}
This reweighting requires assigning parton-shower histories (of successive branchings at $t_1$, $t_2$ and $t_3$) to the $+3$-parton tree-level phase space points. To guarantee that the all-order description of the shower is not deteriorated, it is important to consider all possible parton shower histories, as discussed at length in \cite{Prestel:2021vww}. The reweighting is identical to that within the \textsc{Ckkw-l} tree-level merging scheme~\cite{Lonnblad:2001iq,Lonnblad:2011xx}. After the reweighting, the contribution produces, due to the inclusion of no-emission probabilities, a physically meaningful three-jet spectrum, even when partons become (successively) unresolved. As such, it is an appropriate description of the real-emission corrections to $\Phi_{n+2}$ states.

The \nlo{} cross section for $\Phi_{n+2}$ states consists of tree-level, virtual and real-emission contributions. The latter can be separated into a single-parton unresolved piece (defined here as configurations for which no parton-shower history with three emissions above the shower cut-off $\sim 1$ GeV can be found) and a resolved component. The former is combined with the virtual corrections into the ``exclusive" first-order correction to the $\Phi_{n+2}$ rate. This exclusive correction needs to be combined with the resolved real-emission component to obtain the full inclusive \nlo{} correction -- whose rate is naively given by the three-parton contributions before reweighting discussed above. 

To serve as suitable (\nnnlopps{}) description of $\Phi_{n+2}$ states, any rate has to incorporate the no-emission probabilities to allow partons to become unresolved, and include the effect of the (dynamical) renormalization- and factorization-scale setting mechanism of the parton shower. Thus, the
exclusive first-order correction to the $\Phi_{n+2}$ is amended with shower all-order factors:
\begin{equation*}
\textnormal{} 
\quad
\vcenter{\hbox{

}}
\quad .
\end{equation*}
This is the cross section that needs to be complemented with resolved three-parton configurations. At the same time, the three-parton spectrum discussed above should not lead to double-counting, and thus needs to be removed from the inclusive rate. Overall, this means that the contribution to $\Phi_{n+2}$ states stemming from fixed-order terms of $\mathcal{O}(\alpha_s^3)$ is given by

\begin{equation*}
\vcenter{\hbox{
%
}}
\end{equation*}
where the second term complements the exclusive cross section, and the third
term acts as all-order subtraction of the (reweighted) fully differential
three-parton rate.

\begin{figure}[tbp]
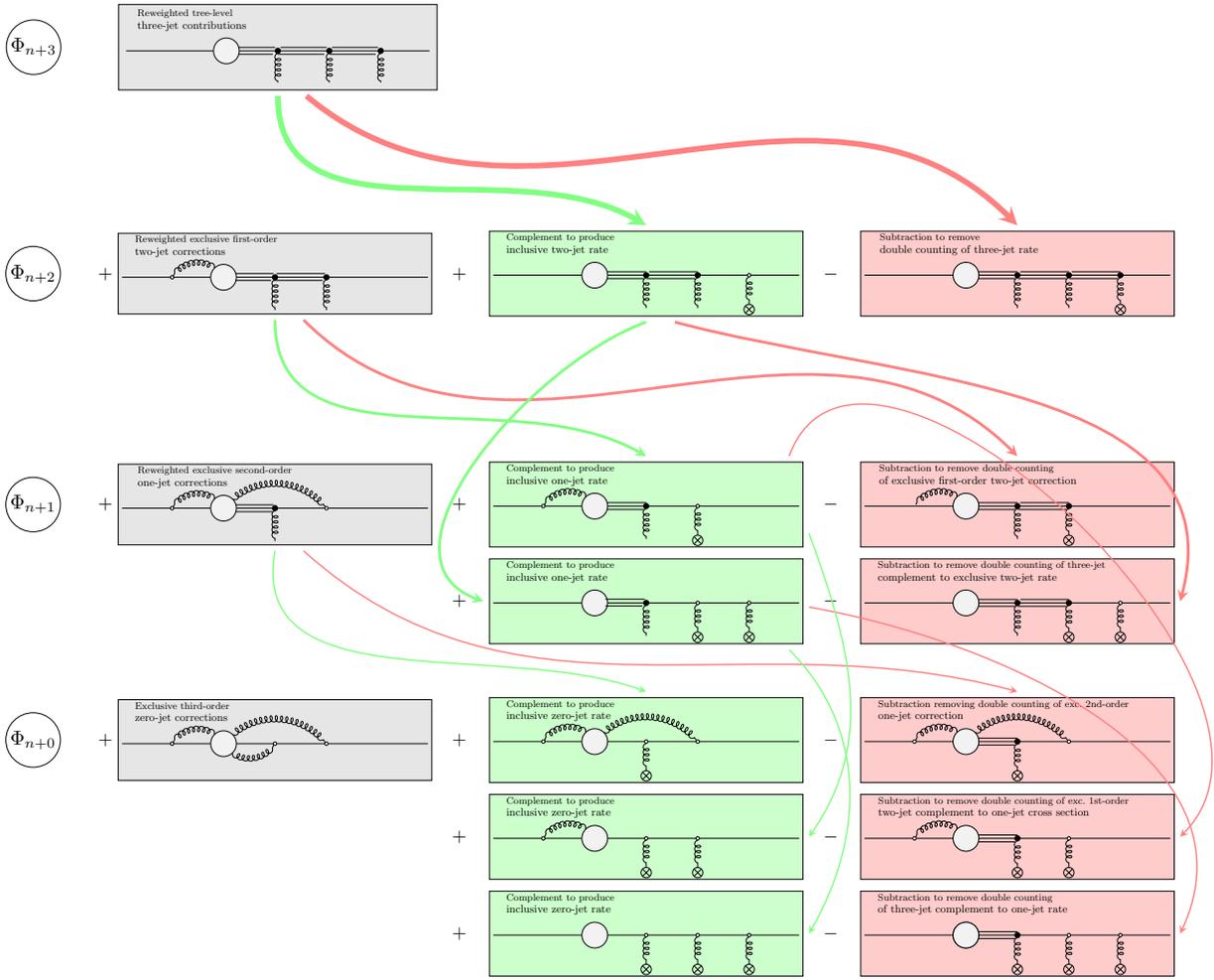

\centering
\resizebox{1.0\textwidth}{!}{
%
};

\node (minus) [nobox,left=of lops321sub.west,xshift=0.65cm]{$-$};

\draw[style={font=\sffamily\small}]
    ($(lops3.south)+(0.0,-0.1)$) edge [->,>=stealth,out=-90,in=150,green!50,line width=1mm] ($(lops3com.north)+(0.0,0.1)$);
\draw[style={font=\sffamily\small}]
    ($(lops3.south)+(0.5,-0.1)$) edge [->,>=stealth,out=-40,in=135,red!50,line width=1mm] ($(lops3sub.north)+(0.0,0.1)$);
\draw[style={font=\sffamily\small}]
    ($(excnlops2.south)+(0.0,-0.1)$) edge [->,>=stealth,out=-90,in=150,green!50,line width=0.5mm] ($(excnlops2com.north)+(0.0,0.1)$);
\draw[style={font=\sffamily\small}]
    ($(excnlops2.south)+(0.5,-0.1)$) edge [->,>=stealth,out=-45,in=135,red!50,line width=0.5mm] ($(excnlops2sub.north)+(0.0,0.1)$);
\draw[style={font=\sffamily\small}]
    ($(lops3com.south)+(0.0,-0.1)$) edge [->,>=stealth,out=-160,in=160,green!50,line width=0.5mm] ($(lops32com.west)+(-0.1,0.0)$);
\draw[style={font=\sffamily\small}]
    ($(lops3com.south)+(0.5,-0.1)$) edge [->,>=stealth,out=-15,in=75,red!50,line width=0.5mm] ($(lops32sub.east)+(0.1,0.0)$);
\draw[style={font=\sffamily\small}]
    ($(excnlops1.south)+(0.0,-0.1)$) edge [->,>=stealth,out=-100,in=160,green!50,line width=0.25mm] ($(excnlops1com.north)+(0.0,0.1)$);
\draw[style={font=\sffamily\small}]
    ($(excnlops1.south)+(0.5,-0.1)$) edge [->,>=stealth,out=-42,in=165,red!50,line width=0.25mm] ($(excnlops1sub.north)+(0.0,0.1)$);
\draw[style={font=\sffamily\small}]
    ($(lops32com.south)+(2.5,-0.1)$) edge [->,>=stealth,out=-45,in=60,green!50,line width=0.25mm] ($(lops321com.east)+(0.1,0.0)$);
\draw[style={font=\sffamily\small}]
    ($(lops32com.east)+(0.1,-0.1)$) edge [->,>=stealth,out=-10,in=65,red!50,line width=0.25mm] ($(lops321sub.east)+(0.1,0.0)$);
\draw[style={font=\sffamily\small}]
    ($(excnlops2com.east)+(0.1,-0.5)$) edge [->,>=stealth,out=-70,in=55,green!50,line width=0.25mm] ($(excnlops21com.east)+(0.1,0.0)$);
\draw[style={font=\sffamily\small}]
    ($(excnlops2com.north)+(2.5,0.1)$) edge [->,>=stealth,out=70,in=45,red!50,line width=0.25mm] ($(excnlops21sub.east)+(0.1,0.0)$);

\end{tikzpicture}
}
\caption{\label{fig:oas3-contributions} Diagrammatic representation of the $\mathcal{O}(\alpha_s^{n+3})$  contributions to the \nnnlops{} method. Grey boxes contain (all-order reweighted) fixed-order samples, green boxes complements for exclusive cross sections, and red boxes all-order subtractions to ensure unitarity.
The multiplicity of contributions decreases from top to bottom, with the top layer containing $\Phi_{n+3}$ phase space points, the next-lower layer $\Phi_{n+2}$, followed by $\Phi_{n+1}$ and finally $\Phi_{n}$ phase space points in the bottom layer. Arrows have been added to highlight the relations and/or cancellation between different contributions.}
\end{figure}

This discussion highlights the core construction principles of the \nnnlops{} method using exclusive cross sections:
\begin{itemize}
\item[$a)$] Create and add a physically meaningful real-emission pattern for $m$-parton states, by reweighting fixed-order results with appropriate shower all-order factors
\item[$b)$] Subtract the real-emission pattern that has been added from the next-lower multiplicity $(m-1)$
\item[$c)$] Create and add a physically meaningful, and appropriate, higher-order exclusive cross section for the $(m-1)$-parton states
\item[$d)$] Complement the exclusive $(m-1)$-parton cross section with an identically-weighted resolved $m$-parton real-emission pattern; an unbiased inclusive cross section requires the introduction of ``bias correction factors" $\one{m}{n}$~\cite{Prestel:2021vww}.
\end{itemize}
This yields an $(m-1)$-parton matched calculation that may serve as ``physically meaningful real-emission pattern" for the next-lower multiplicity $(m-2)$, and which may be used to iterate the procedure from $b)$ onwards, until the minimal multiplicity is reached. The full procedure for contributions with a fixed-order
expansion starting at $\mathcal{O}(\alpha_s^3)$ is shown in Figure~\ref{fig:oas3-contributions}.

Reweighting any contribution starting at $\mathcal{O}(\alpha_s^3)$ will not result change fixed-order coefficients at or below $\mathcal{O}(\alpha_s^3)$, so that the contributions shown in Figure~\ref{fig:oas3-contributions} both preserve the precision of the fixed-order prediction and the accuracy of the all-order shower prescription.
Any reweighting of contribution starting at $\mathcal{O}(\alpha_s^1)$ or $\mathcal{O}(\alpha_s^2)$ will, however, naively lead to problematic higher-order terms. The reweighting should be appropriately subtracted to avoid this issue. The $\mathcal{O}(\alpha_s^1)$-term of the no-emission probability will be depicted by  
\begin{equation*}
\textnormal{} 
\quad
\vcenter{\hbox{
%
}}
\quad
\textnormal{,} 
\end{equation*}
while the $\mathcal{O}(\alpha_s^1)$-term of the shower vertex is illustrated by
\begin{equation*}
\textnormal{} 
\quad
\vcenter{\hbox{
%
%
}}
\quad
\textnormal{.} 
\end{equation*}
For ease of illustration, it is useful to combine these terms with the all-order no-emission probability and shower vertex to produce an $\mathcal{O}(\alpha_s^1)$-subtracted parton shower weight. For this, we interpret the product of all all-order factors as a unique weight with a single, common, expansion, and introduce the subtracted weight of $\Phi_{n+2}$ states by
\begin{eqnarray*}
\vcenter{\hbox{
%
%
}}
~
\Bigg]~.
\end{eqnarray*}
The $\mathcal{O}(\alpha_s^1)$-subtracted parton shower weight for $\Phi_{n+1}$ states
is defined analogously as
\begin{eqnarray}
\label{eq:ps1_oas1_subtracted}
\vcenter{\hbox{
%
%
}}
~\Bigg]\nonumber
\end{eqnarray}
When reweighting contributions starting at $\mathcal{O}(\alpha_s^1)$, the weight should not contain any $\mathcal{O}(\alpha_s^2)$ terms. Thus, we introduce the diagram
\begin{equation*}
\textnormal{} 
\quad
\vcenter{\hbox{
%
%
}}
\quad
\textnormal{,} 
\end{equation*}
to describe the $\mathcal{O}(\alpha_s^2)$-term of the no-emission probability, 
while the $\mathcal{O}(\alpha_s^2)$-term of the shower vertex is given by
\begin{equation*}
\textnormal{} 
\quad
\vcenter{\hbox{
%
%
}}
\quad
\textnormal{.} 
\end{equation*}
With this, the The $\mathcal{O}(\alpha_s^2)$-subtracted parton shower weight for $\Phi_{n+1}$ states
is
\begin{eqnarray*}
\vcenter{\hbox{
%
%
}}
~+~
\textnormal{corrections due to eq.~\ref{eq:ps1_oas1_subtracted}}
~\Bigg]
\end{eqnarray*}
where the ``corrections due to eq.~\ref{eq:ps1_oas1_subtracted}" remove undesirable $\mathcal{O}(\alpha_s^2)$ terms that are newly introduced by the $\mathcal{O}(\alpha_s^1)$-subtracted weight defined in eq.~\ref{eq:ps1_oas1_subtracted}. Details on the second-order subtracted weight may be found in Appendix \ref{sec:problemstatement}.

\begin{figure}[h]
\centering
\resizebox{1.0\textwidth}{!}{
%
};

\node (minus) [nobox,left=of lops21sub.west,xshift=0.65cm]{$-$};

\draw[style={font=\sffamily\small}]
    ($(lops2.south)+(0.0,-0.1)$) edge [->,>=stealth,out=-90,in=150,green!50,line width=1mm] ($(lops2com.north)+(0.0,0.1)$);
\draw[style={font=\sffamily\small}]
    ($(lops2.south)+(0.5,-0.1)$) edge [->,>=stealth,out=-45,in=135,red!50,line width=1mm] ($(lops2sub.north)+(0.0,0.1)$);
\draw[style={font=\sffamily\small}]
    ($(lops2com.south)+(0.0,-0.1)$) edge [->,>=stealth,out=-160,in=160,green!50,line width=0.5mm] ($(lops21com.west)+(-0.1,0.0)$);
\draw[style={font=\sffamily\small}]
    ($(lops2com.south)+(0.5,-0.1)$) edge [->,>=stealth,out=-15,in=75,red!50,line width=0.5mm] ($(lops21sub.east)+(0.1,0.0)$);
\draw[style={font=\sffamily\small}]
    ($(excnlops1.south)+(0.0,-0.1)$) edge [->,>=stealth,out=-100,in=160,green!50,line width=0.5mm] ($(excnlops1com.north)+(0.0,0.1)$);
\draw[style={font=\sffamily\small}]
    ($(excnlops1.south)+(0.5,-0.1)$) edge [->,>=stealth,out=-42,in=165,red!50,line width=0.5mm] ($(excnlops1sub.north)+(0.0,0.1)$);

\end{tikzpicture}
}
\caption{\label{fig:oas2-contributions} Diagrammatic representation of the $\mathcal{O}(\alpha_s^{n+2})$  contributions to the \nnnlops{} method. The color coding is identical to that in Figure~\ref{fig:oas2-contributions}, and as before, the multiplicity of contributions decreases from top to bottom. Arrows have been added to highlight the relations/cancellation between different contributions.}
\end{figure}

\begin{figure}[h]
\centering
\resizebox{1.0\textwidth}{!}{
\begin{tikzpicture}[remember picture]
\node (lops1) [box, xshift=0.0cm, scale=0.6,fill=mymidgrey!20] {
\begin{tikzpicture}
  \begin{feynman} [every blob={/tikz/fill=gray!10,/tikz/inner sep=2pt}]
    \vertex [xshift=-0.5cm] (a) {\smaller };
    \vertex [right=of a] (v0) {\smaller };
    \vertex [right=of v0, blob] (b) {\smaller };
    \vertex [] (bu) at ($(b)+(0.31,0.1)$) {\smaller } ;
    \vertex [] (bd) at ($(b)+(0.31,-0.1)$) {\smaller };
    %\vertex [dot,right=of b, xshift=0.0cm] (v1) {\smaller };
    \vertex [circ, scale=2.0,right= of b, xshift=0.0cm,/tikz/fill=red!20] (v1x) {\smaller};
    \vertex [crossed dot, scale=0.45,right=of b, xshift=0.0cm,/tikz/fill=gray!20] (v1) {\smaller };
    \vertex [below=of v1, yshift=0.5cm] (glu1) {\smaller };
    \vertex [] (v1u) at ($(v1)+(0.0,0.1)$) {\smaller };
    \vertex [] (v1d) at ($(v1)+(-0.03,-0.1)$) {\smaller };
    \vertex [right=of v1] (c) {\smaller };
    \vertex [above=of b,yshift=-0.6cm,xshift=-1.2cm,align=left] (title) {\smaller Reweighted tree-level\\ one-jet corrections};
    \diagram* {
      (a) -- (b),
      (b) -- [insertion=0.4,insertion=0.6] (v1x),
      (b) -- (v1),
      (v1) -- (c),
      (bu) -- (v1u),
      (bd) -- (v1d),
      (v1) -- [gluon] (glu1)
    };
  \end{feynman}
\end{tikzpicture}
};

\node (phi1) [circ,left=of lops1.west,xshift=0.0cm]{\larger $\Phi_{n+1}$};

\node (excnlops0) [box, below=of lops1.south, yshift=-1.62cm, scale=0.6,fill=mymidgrey!20] {
\begin{tikzpicture}
  \begin{feynman} [every blob={/tikz/fill=gray!10,/tikz/inner sep=2pt}]
    \vertex [xshift=-0.5cm] (a) {\smaller };
    \vertex [circ,right=of a] (v0) {\smaller };
    \vertex [right=of v0, blob] (b) {\smaller };
    \vertex [] (bu) at ($(b)+(0.27,0.1)$) {\smaller } ;
    \vertex [] (bd) at ($(b)+(0.27,-0.1)$) {\smaller };
    \vertex [right=of b, xshift=0.0cm] (v1) {\smaller };
    \vertex [below=of v1, yshift=0.5cm] (glu1) {\smaller };
    \vertex [] (v1u) at ($(v1)+(0.0,0.1)$) {\smaller };
    \vertex [] (v1d) at ($(v1)+(-0.07,-0.1)$) {\smaller };
    \vertex [right=of v1] (c) {\smaller };
    \vertex [above=of b,yshift=-0.6cm,xshift=-1.2cm,align=left] (title) {\smaller Exclusive first-order\\ zero-jet corrections};
    \diagram* {
      (a) -- (v0),
      (v0) -- (b),
      (b) -- (c),
      (b) -- [gluon, quarter right] (v0)
    };
  \end{feynman}
\end{tikzpicture}
};

\node (plus) [nobox,left=of excnlops0.west,xshift=0.95cm]{+};

\node (lops1com) [box, right=of excnlops0, scale=0.6,fill=green!20] {
\begin{tikzpicture}
  \begin{feynman} [every blob={/tikz/fill=gray!10,/tikz/inner sep=2pt}]
    \vertex [xshift=-0.5cm] (a) {\smaller };
    \vertex [right=of a] (v0) {\smaller };
    \vertex [right=of v0, blob] (b) {\smaller };
    \vertex [] (bu) at ($(b)+(0.27,0.1)$) {\smaller } ;
    \vertex [] (bd) at ($(b)+(0.27,-0.1)$) {\smaller };
    \vertex [circ,right=of b, xshift=0.0cm] (v1) {\smaller };
    \vertex [crossed dot, below=of v1, yshift=0.5cm] (glu1) {\smaller };
    \vertex [] (v1u) at ($(v1)+(0.0,0.1)$) {\smaller };
    \vertex [] (v1d) at ($(v1)+(-0.07,-0.1)$) {\smaller };
    \vertex [right=of v1] (c) {\smaller };
    \vertex [above=of b,yshift=-0.6cm,xshift=-1.0cm,align=left] (title) {\smaller Complement to produce\\ inclusive zero-jet rate};
    \diagram* {
      (a) -- (b),
      (b) -- (v1),
      (v1) -- (c),
      (v1) -- [gluon] (glu1)
    };
  \end{feynman}
\end{tikzpicture}
};

\node (plus) [nobox,left=of lops1com.west,xshift=0.65cm]{+};

\node (phi0) [circ,left=of excnlops0.west,xshift=0.0cm]{\larger $\Phi_{n+0}$};

\node (lops1sub) [box, right=of lops1com,scale=0.6,fill=red!20] {
\begin{tikzpicture}
  \begin{feynman} [every blob={/tikz/fill=gray!10,/tikz/inner sep=2pt}]
    \vertex [xshift=-0.5cm] (a) {\smaller };
    \vertex [right=of a] (v0) {\smaller };
    \vertex [right=of v0, blob] (b) {\smaller };
    \vertex [] (bu) at ($(b)+(0.27,0.1)$) {\smaller } ;
    \vertex [] (bd) at ($(b)+(0.27,-0.1)$) {\smaller };
    \vertex [circ, scale=2.0,right= of b, xshift=0.0cm,/tikz/fill=red!20] (v1x) {\smaller};
    \vertex [crossed dot, scale=0.45,right=of b, xshift=0.0cm,/tikz/fill=gray!20] (v1) {\smaller };
    \vertex [crossed dot, below=of v1, yshift=0.5cm] (glu1) {\smaller };
    \vertex [] (v1u) at ($(v1)+(0.0,0.1)$) {\smaller };
    \vertex [] (v1d) at ($(v1)+(-0.07,-0.1)$) {\smaller };
    \vertex [right=of v1] (c) {\smaller };
    \vertex [above=of b,yshift=-0.6cm,xshift=0.0cm,align=left] (title) {\smaller Subtraction removing double counting\\ of tree-level one-jet correction};
    \diagram* {
      (a) -- (b),
      (b) -- [insertion=0.4,insertion=0.6] (v1x),
      (b) -- (v1),
      (v1) -- (c),
      (bu) -- (v1u),
      (bd) -- (v1d),
      (v1) -- [gluon] (glu1)
    };
  \end{feynman}
\end{tikzpicture}
};

\node (minus) [nobox,left=of lops1sub.west,xshift=0.65cm]{$-$};

\draw[style={font=\sffamily\small}]
    ($(lops1.south)+(0.0,-0.1)$) edge [->,>=stealth,out=-90,in=150,green!50,line width=1mm] ($(lops1com.north)+(0.0,0.1)$);
\draw[style={font=\sffamily\small}]
    ($(lops1.south)+(0.5,-0.1)$) edge [->,>=stealth,out=-45,in=135,red!50,line width=1mm] ($(lops1sub.north)+(0.0,0.1)$);

\end{tikzpicture}
}
\caption{\label{fig:oas1-contributions} Diagrammatic representation of the $\mathcal{O}(\alpha_s^{n+1})$  contributions to the \nnnlops{} method. The color coding is identical to that in Figure~\ref{fig:oas2-contributions}, and as before, the multiplicity of contributions decreases from top to bottom. Arrows have been added to highlight the relations/cancellation between different contributions.}
\end{figure}
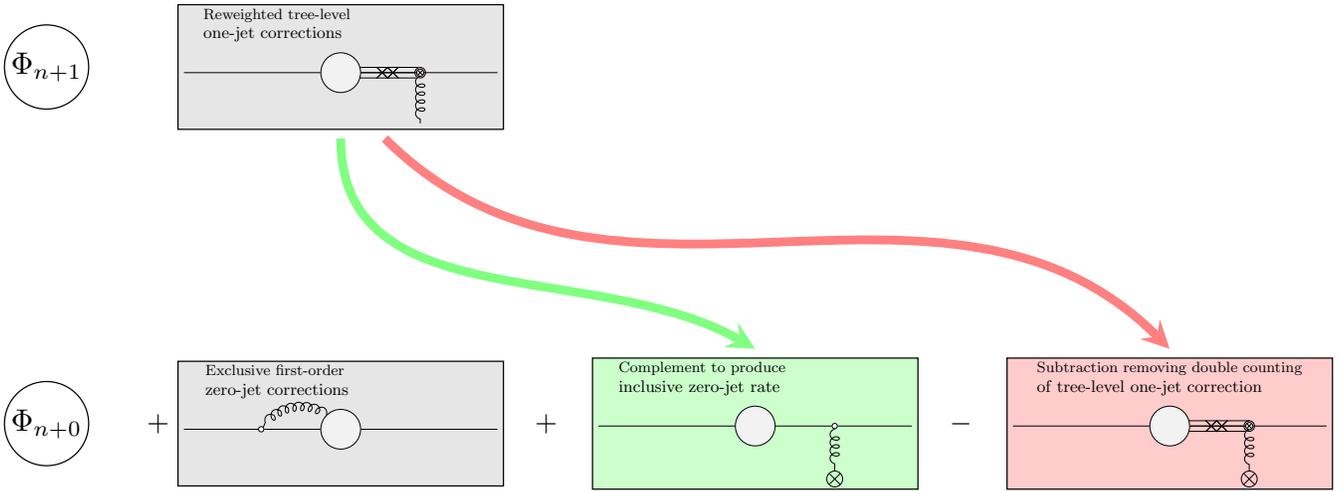

Using these definitions, the remaining parts of the \nnnlops{} scheme may be constructed. Figure~\ref{fig:oas2-contributions} shows the treatment of contributions starting at $\mathcal{O}(\alpha_s^2)$, while Figure~\ref{fig:oas1-contributions} illustrates the treatment of terms starting at $\mathcal{O}(\alpha_s^1)$. The \nnnlops{} method then stipulates that
\begin{equation*}
\langle \obs{}\rangle_{\nnnlops{}}
~=
~
\vcenter{\hbox{
\begin{tikzpicture}[remember picture]
\node (def) [nobox,inner sep=2pt, inner ysep=2pt,scale=0.6] {
\begin{tikzpicture}%
  \begin{feynman} [every blob={/tikz/fill=gray!10,/tikz/inner sep=2pt}]%
    \vertex [xshift=0.0cm] (a) {\smaller };%
    \vertex [right=of a, blob] (b) {\smaller };%
    %\vertex [below=of b, yshift=0.5cm] (glu1) {\smaller \phantom{glu1}};
    \vertex [right=of b] (c) {\smaller };%
    %\vertex [above=of b,yshift=-0.6cm, align=center] (title) {\smaller Hard process};
    \diagram* { (a) -- (b), (b) -- (c); };%
  \end{feynman}%
\end{tikzpicture}%
};
\end{tikzpicture}%
}}
~+~
\mathrm{Figure}~\ref{fig:oas1-contributions}
~+~
\mathrm{Figure}~\ref{fig:oas2-contributions}
~+~
\mathrm{Figure}~\ref{fig:oas3-contributions}
\end{equation*}
leads to \nnnlopps{}-correct predictions. The same matching formula in more mathematical detail is given by eq.\ $(24)$ of \cite{Prestel:2021vww}.

%\begin{itemize}
%\item shower as an add-subtract scheme
%\item show shower up to third emission
%\item goals of N3LO matching
%\item explain going from high to low multiplicity
%\item argue how to replace fixed-order coefficients in the shower three-emission
%expression. link to ckkwl.
%\item nlo matching of two-jet rate with unlops, and exclusive cross sections, starting from previous 
%expression. apply all sudakov factors
%\item nnlo matching of one-jet rate, as discussed in tomte paper.
%\item n3lo+ps
%\end{itemize}

\subsection{Complications for hadronic initial states}

When applying the \nnnlops{} method to hadronic collisions, several complications arise relative to its application to leptonic initial states. The previous section introduced the shower all-order weights and their expansions as abstract objects. More concretely, these weights incorporate the effects of $a)$ resummation of unresolved emissions for states with large scale hierarchies by means of no-emission probabilities, $b)$ dynamical renormalization scale setting for multi-jet processes and $c)$ dynamical factorization scale setting in the presence of incoming identified hadrons. Points $a)$ and $b)$ do already emerge in the discussion of the evolution of $e^+e^-\rightarrow$ jets scattering, and have been discussed in \cite{Prestel:2021vww}. However, due to to the unitarity of the parton-shower evolution, points $a)$ and $c)$ are linked: The no-emission factors providing the resummation of large scale hierarchies in $a)$ contain information about factorization scales, and the about parton distributions. Thus,
\begin{itemize}
\item An appropriate reweighting of emission rates includes ratios of \pdf{}s to produce the all-order effects of dynamical factorization scale setting;
\item No-emission probabilities will contain exponentiated \pdf{} ratios. 
\end{itemize}
The no-emission probability in the presence of initial-state partons is given by
\begin{equation}
\textnormal{} 
\quad
\vcenter{\hbox{
\begin{tikzpicture}[remember picture]
\node (def) [nobox,inner sep=2pt, inner ysep=2pt,scale=0.7] {
\begin{tikzpicture}
  \begin{feynman} [every blob={/tikz/fill=gray!10,/tikz/inner sep=2pt}]
    \vertex [] (a) {\smaller };
    \vertex [circ, right=of a, xshift=0.0cm] (v1) {\smaller };
    \vertex [below=of v1, yshift=0.5cm, align=center] (glu1) {\smaller $t$};
    \vertex [circ, right=of v1, xshift=0.0cm] (v2) {\smaller };
    \vertex [below=of v2, yshift=0.5cm, align=center] (glu2) {\smaller $\bar t$};
    \vertex [right=of a, yshift=0.1cm] (v1u)  {\smaller };
    \vertex [right=of a, yshift=-0.1cm] (v1d) {\smaller };
    \vertex [right=of v1u] (v2u) {\smaller };
    \vertex [right=of v1d] (v2d) {\smaller };
    \vertex [right=of v2] (c) {\smaller };
    %\vertex [above=of v1,yshift=-0.6cm,xshift=0.8cm, align=center] (title) {\smaller No emission between $t_1$ and $t_2$ };
    \diagram* {
      (a) -- (v1),
      (v1) -- (v2),
      (v2) -- (c),
      (v1u) -- (v2u),
      (v1d) -- (v2d),
      (v1) -- [scalar] (glu1),
      (v2) -- [scalar] (glu2)
    };
  \end{feynman}
\end{tikzpicture}
};
\end{tikzpicture}%
}}
~=~
\prod_{r_i\in\Phi_n}
\Pi_{r_i}(x_i(\Phi_n); t, \bar t)
\end{equation}
where the product runs over all distinct (sets of) particles $r_i$ that may emit radiation (e.g.\ all radiating dipole ends in a partial-fractioned dipole shower like~\cite{Sjostrand:2004ef,Schumann:2007mg,Platzer:2009jq,Hoche:2015sya}), and the ``single-radiator" no-emission probabilities are defined by
\begin{eqnarray}
&&-\ln\left\{
\Pi_{r_i}(x_i(\Phi_n); t, \bar t)
\right\}\\
&&
\quad
=
\begin{cases}
  \int\limits_{\bar t}^{t} \frac{d\rho}{\rho} \sum\limits_{s(r_i)} \int\limits_{\Omega(s,\rho)} dz \frac{\alpha_s(\kappa)}{2\pi}
\pskernel{f}{f'(s)}{r_i,\Phi_n, \Phi_{n+1}}
 & \qquad\mathbf{A}\\
  \int\limits_{\bar t}^{t} \frac{d\rho}{\rho} \sum\limits_{s(r_i)} \int\limits_{\Omega(s,\rho)} dz \frac{\alpha_s(\kappa)}{2\pi}
\frac{\xf{p'(s)}{x(p'(s),\Phi_{n+1})}{\rho}}{\xf{p(s)}{x(p(s),\Phi_n)}{\rho}} 
\pskernel{f}{f'(s)}{r_i,\Phi_n, \Phi_{n+1}}
 & \qquad\mathbf{B}
\end{cases}\nonumber
\end{eqnarray}
for the splittings ``$s(r_i)$" only affecting final-state particles (case {\bf A}), or also affecting the initial-state particle $p(s)$ by changing it to a post-branching particle $p'(s)$ (case {\bf B}), and where $\Phi_{n+1}$ is the union of $\Phi_{n}$ and the shower phase space sampling variables $(z,\rho,\phi)$, and where $\pskernel{f}{f'(s)}{r_i,\Phi_n, \Phi_{n+1}}$ is the shower splitting kernel for the splitting $s(r_i)$. The renormalization scale $\kappa$ may depend on the splitting. The phase space boundaries $\Omega$ are derived purely from momentum conservation in case {\bf A}, while for case {\bf B}, the constraint $x(\Phi_{n+1})>x(\Phi_{n})$ due to backward initial-state evolution ~\cite{Sjostrand:1985xi} enters additionally.

For splittings $s(r_i)$ affecting initial-state particles of a pre-branching state $\Phi_n$, the emission vertex is given by
\begin{equation*}
\textnormal{} 
\quad
\vcenter{\hbox{
\begin{tikzpicture}[remember picture]
\node (def) [nobox,inner sep=2pt, inner ysep=2pt,scale=0.7] {
\begin{tikzpicture}
  \begin{feynman} [every blob={/tikz/fill=gray!10,/tikz/inner sep=2pt}]
    \vertex [xshift=0.0cm] (a) {\smaller };
    \vertex [dot, right=of a] (v1) {\smaller };
    %\vertex [below=of v1, yshift=0.5cm, align=center] (glu1) {\smaller soft/coll.\ emission at $t$};
    \vertex [below=of v1, yshift=0.5cm, align=center] (glu1) {\smaller $t$};
    \vertex [right=of v1] (c) {\smaller };
    %\vertex [above=of v1,yshift=-0.6cm, align=center] (title) {\smaller Shower vertex at $t$ };
    \diagram* {
      (a) -- (v1),
      (v1) -- (c),
      (v1) -- [scalar] (glu1)
    };
  \end{feynman}
\end{tikzpicture}
};
\end{tikzpicture}%
}}
~=~
\frac{\alpha_s(\kappa)}{\alpha_s(\mu_\mathrm{R})}
\frac{\xf{p'(s)}{x(p'(s),\Phi_{n+1})}{t}}{\xf{p(s)}{x(p(s),\Phi_n)}{t}} 
\textnormal{.} 
\end{equation*}
The product of no-emission probabilities and shower vertices provides a suitable all-order reweighting of fixed-order calculations, as e.g.\ realized in~\cite{Lonnblad:2001iq,Lonnblad:2011xx} within the context of leading-order merging prescriptions. The product yields the effect of Sudakov resummation when jets become individually unresolved\footnote{Common transverse-momentum ordered parton showers aim to at least describe leading logarithms for observables linearly related to their evolution variable. Since parton showers include several (semi-)universal effects beyond leading logarithm, the description is, in practise, superior to lowest-order analytic calculations. An accurate log-counting for the parton shower result of specific observables is, however, challenging due to the use of exact kinematics.}. 

Thus, we will apply the all-order shower weight 
\begin{eqnarray}
\label{eq:ckkwl-weight}
w_n &=& 
\frac{\xf{n+}{x_n^+}{t_{n}} }{ \xf{n+}{x_n^+}{\mu_\mathrm{F}} }
\frac{\xf{n-}{x_n^-}{t_{n}}}{\xf{n-}{x_n^-}{\mu_\mathrm{F}}}
\prod_{i=1}^n\left(
\frac{\xf{0+}{x_0^+}{t_{i-1}} }{ \xf{0+}{x_0^+}{t_i} }
\frac{\xf{0-}{x_0^-}{t_{i-1}}}{\xf{0-}{x_0^-}{t_i}}
\frac{\alpha_s(t_i)}{\alpha_s(\mu_\mathrm{R})}
\right)
\Pi_{r_i}(x_i(\Phi_n); t_{i-1}, t_i)~.
\end{eqnarray}
to all fixed-order input for $\Phi_n$, with one or more final-state partons. The $\pm$ indexes the incoming hadrons in with large $p^\pm$-momentum, and $t_0=\mu_\mathrm{F}$ is used. Whenever the application of this weight induces undesirable behavior at or below $\mathcal{O}(\alpha_s^3)$, the unwanted coefficients in its expansion are removed (by subtraction) to ensure appropriate behavior, as discussed in sec.~\ref{sec:basics}.
The presence of \pdf{} ratios makes the order-by-order expansion of the weight cumbersome. The $\mathcal{O}(\alpha_s)$-coefficient of the expansion of such weights is a necessary ingredient in the \unlops{} \nlo{} merging scheme, and is documented in~\cite{Lonnblad:2012ix}. However, for \nnnlopps{} matching, the $\mathcal{O}(\alpha_s^2)$ expansion of this weight is required. The result is somewhat lengthy, and comprises the major complication when applying the \nnnlops{} method to hadronic collisions. All necessary ingredients are presented in Appendix \ref{sec:second-order-expansion}.

%\begin{itemize}
%\item The (PDF ratio) weight assigned to contributions that multiply bias correction factors is not CKKW-L-like (since it has to remove information on higher-multiplicity PDFs).
%\item The latter might be conveniently treated by absorbing the relevant PDF ratios into the bias correction factors.
%\item treatment of MPI
%\end{itemize}

It is important to note that in order to complement the fixed-order cross sections, it is necessary to apply appropriate \pdf{} factors to reclustered complements. If, for example, a three-parton configuration is employed to complement the exclusive one-parton cross section, then the fact that the former has been pre-tabulated with \pdf{}s depending on the initial partons entering in $\Phi_{n+3}$ at factorization scale $\mu_\mathrm{F}$ has to be reflected in the weight. This implies the change
\begin{eqnarray}
\frac{\xf{n+}{x_n^+}{t_{n}} }{ \xf{n+}{x_n^+}{\mu_\mathrm{F}} }
\frac{\xf{n-}{x_n^-}{t_{n}}}{\xf{n-}{x_n^-}{\mu_\mathrm{F}}}
\rightarrow
\frac{\xf{n+}{x_n^+}{t_{n}} }{ \xf{m+}{x_m^+}{\mu_\mathrm{F}} }
\frac{\xf{n-}{x_n^-}{t_{n}}}{\xf{m-}{x_m^-}{\mu_\mathrm{F}}}\quad \textnormal{($m\geq n$)},
\end{eqnarray}
in the first factor in the weight defined in eq.~\ref{eq:ckkwl-weight}. Noting that
\begin{eqnarray*}
\frac{\xf{n+}{x_n^+}{t_{n}} }{ \xf{m+}{x_m^+}{\mu_\mathrm{F}} }
\frac{\xf{n-}{x_n^-}{t_{n}}}{\xf{m-}{x_m^-}{\mu_\mathrm{F}}}
&=&
\frac{\xf{n+}{x_n^+}{t_{n}} }{ \xf{n+}{x_n^+}{\mu_\mathrm{F}} }
\frac{\xf{n-}{x_n^-}{t_{n}}}{\xf{n-}{x_n^-}{\mu_\mathrm{F}}}\\
&&\cdot
\underbrace{\left(
\prod_{i=n+1}^m
\frac{\xf{(i-1)+}{x_{i-1}^+}{\mu_\mathrm{F}}} { \xf{n+}{x_n^+}{\mu_\mathrm{F}} }
\frac{\xf{(i-1)-}{x_{i-1}^-}{\mu_\mathrm{F}}} {\xf{n-}{x_n^-}{\mu_\mathrm{F}}}
\right)}_{w_{m\rightarrow n}^{\pdf{}}}
\end{eqnarray*}
offers a simple strategy: The all-order weight applied to complements
is given by the original formulation (eq.~\ref{eq:ckkwl-weight}), as is the case
for the exclusive counterparts, while the \emph{bias-correction factors} $\one{m}{n}$ are used to absorb the rescaling $w_{m\rightarrow n}^{\pdf{}}$.

Finally, high-energy collisions between hadrons feature double- and multiple-parton scattering effects. Naively, the contribution of a secondary \qcd{} interaction arises at $\mathcal{O}(\alpha_s^2)$, and thus warrants a discussion when attempting \nnnlopps{} matching. Assuming an interleaved multiple interaction paradigm~\cite{Sjostrand:1987su,Sjostrand:2004ef} raises the following concerns:
\begin{enumerate}
\item[($1$)] No-secondary-scattering factors $\Pi_\mathrm{MPI}(t_0,t_1)$, containing the probability of no secondary scattering between to scales $t_0>t_1$, need to be considered;
\item[($2$)] Radiation patterns change at $\mathcal{O}(\alpha_s^2)$ and higher, when
adding secondary interactions to the \lo{} and \nlo{} contributions of zero-parton configurations $\Phi_n$; 
\item[($3$)] Hard secondary scatterings change the available longitudinal momentum left in the colliding hadrons, and thus the \pdf{} factors applied to subsequent radiation from the primary hard scattering.
\end{enumerate}
The no-secondary-scattering factors are given by
\begin{equation*}
\Pi_\mathrm{MPI}(t_0,t_1)
=
\exp\left(
-\int\limits_{t_1}^{t_0} \mathrm{d}\sigma_\mathrm{MPI}
\right)
=
1
\,-\,\int\limits_{t_1}^{t_0} \overbrace{\mathrm{d}\sigma_\mathrm{MPI}(\mu_\mathrm{R},\mu_\mathrm{F})}^{\mathcal{O}(\alpha_s^2)}
\,+\,
\mathcal{O}(\alpha_s^3)~,
\end{equation*}
where $\mathrm{d}\sigma_\mathrm{MPI}$ is an appropriately normalized and regularized \qcd{} $2\rightarrow 2$ scattering cross section. Thus, the application of no-secondary-scattering factors introduces 
changes at $\mathcal{O}(\alpha_s^2)$. This implies that if the reweighting of
one-parton contributions includes no-secondary-scattering factors $\Pi_\mathrm{MPI}(t_0,t_1)$, then the tree-level contribution to one-parton states should instead be reweighted by the subtracted no-secondary-scattering factors\footnote{The terms in parentheses can be generated by applying trial shower methods to secondary scattering proposals.}
\begin{equation*}
\Pi_\mathrm{MPI}(t_0,t_1)\,\cdot\,\left(
1
\,-\,\int\limits_{t_1}^{t_0} \mathrm{d}\sigma_\mathrm{MPI}(\mu_\mathrm{R},\mu_\mathrm{F})
\right)~.
\end{equation*}
This shift ensures that only terms of $\mathcal{O}(\alpha_s^4)$ would be introduced by including no-secondary-scattering factors, thus addressing concern ($1$) above.

Concern ($2$) could be handled similarly, since a probability of the first secondary scattering instance occurring at a scale $t_1$ also contains the probability $\Pi_\mathrm{MPI}(t_0,t_1)$ of not producing a harder secondary scattering. This latter probability could be subtracted, thus pushing the impact of secondary scatterings on the radiation pattern to $\mathcal{O}(\alpha_s^4)$. However, also without any subtraction, it is worth noting that the double-radiation states introduced by an \nnlo{} calculation ($2\rightarrow n +2$) have a different structure than those containing a secondary scattering ($2+2\rightarrow n + 2$). If the different states could be disentangled, there would not be any overlap between the calculations, and no subtraction would be necessary. Thus, it appears that concern ($2$) is best investigated on an observable-by-observable basis.

Finally, concern ($3$) first arises for states with three final-state partons. We may suggestively write a parton distribution after a secondary scattering as
\begin{equation*}
f_\mathrm{MPI}(x,\mu)\,=\,
f(x,\mu)
+\Big[\,
f_\mathrm{MPI}(x,\mu)
-
f(x,\mu)
\,\Big] ~.
\end{equation*}
Since secondary scatterings are a higher-order effect, it is reasonable to expect the difference
$[\,
f_\mathrm{MPI}(x,\mu)
-
f(x,\mu)
\,]$ to also be an effect beyond $\mathcal{O}(1)$. This is sufficient to ensure that the change in \pdf{}, which could first manifests for states $\Phi_{n+3}$, would not affect the accuracy of the \nnnlops{} method.

\section{Closure test with Drell-Yan toy N3LO calculation}

\begin{figure}[tbp]
\centering
  \begin{subfigure}{0.47\textwidth}
  \includegraphics[width=0.85\textwidth]{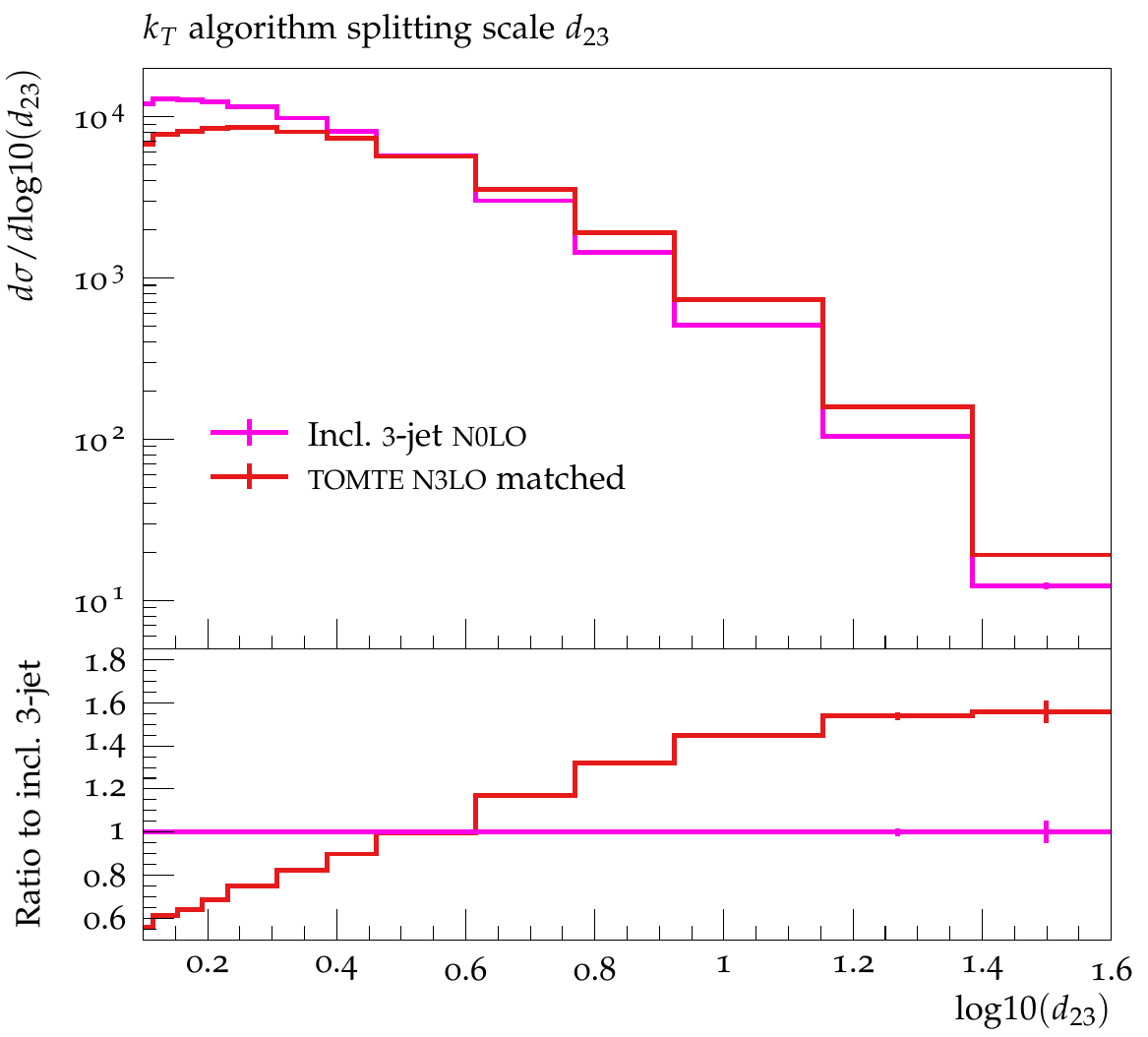}{}
  \caption{\label{fig:tomte-results-dy-3} Separation of the 3rd hardest and 2nd hardest jets, as proxy for (inclusive) observables depending on at least three final-state partons.}
  \end{subfigure}
  \hskip 2mm
  \begin{subfigure}{0.47\textwidth}
  \includegraphics[width=0.85\textwidth]{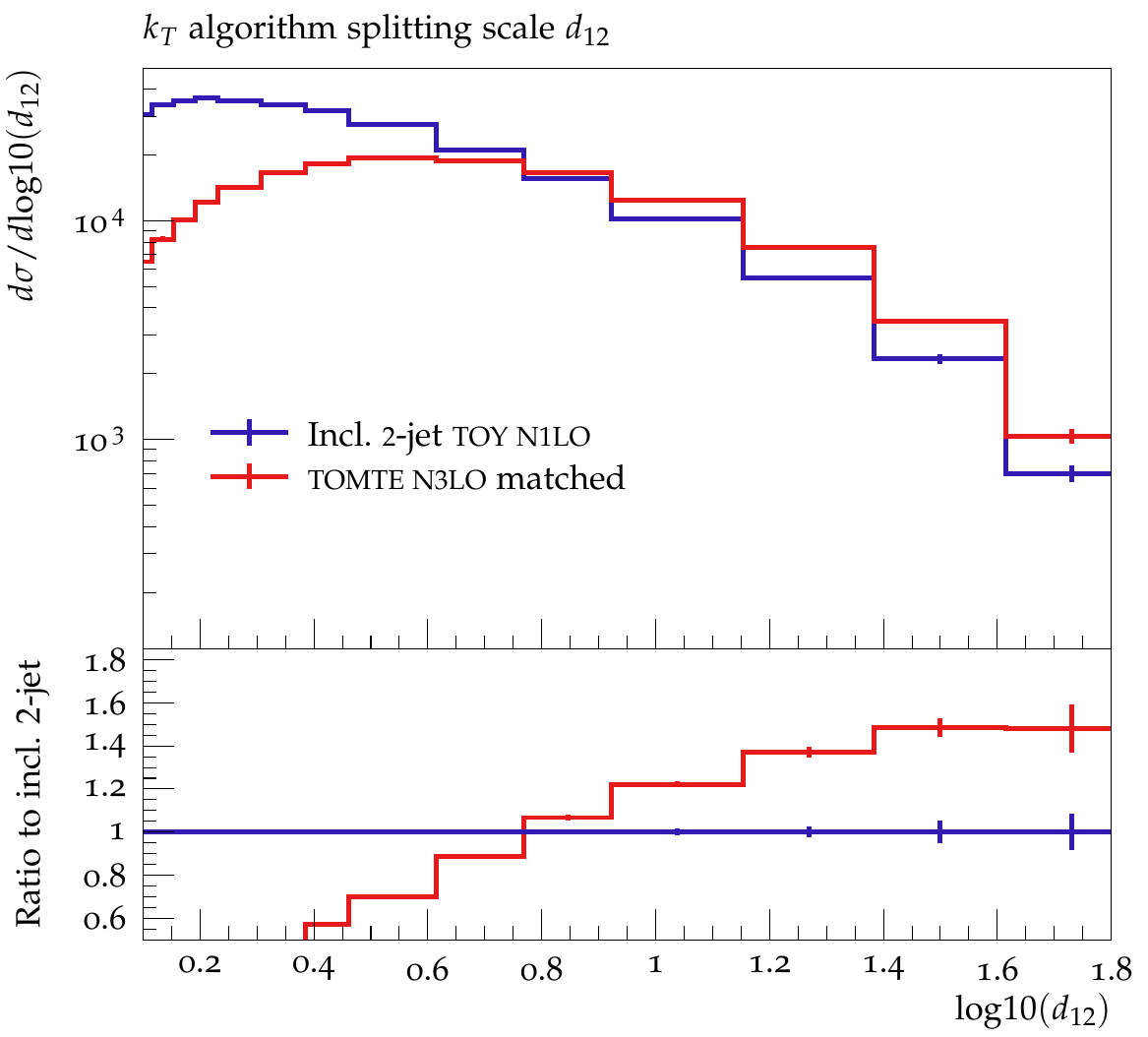}{}
  \caption{\label{fig:tomte-results-dy-2} Separation of the 2nd hardest and hardest jets, as proxy for (inclusive) observables depending on at least two final-state partons.}
  \end{subfigure}\\
  \begin{subfigure}{0.47\textwidth}
  \includegraphics[width=0.85\textwidth]{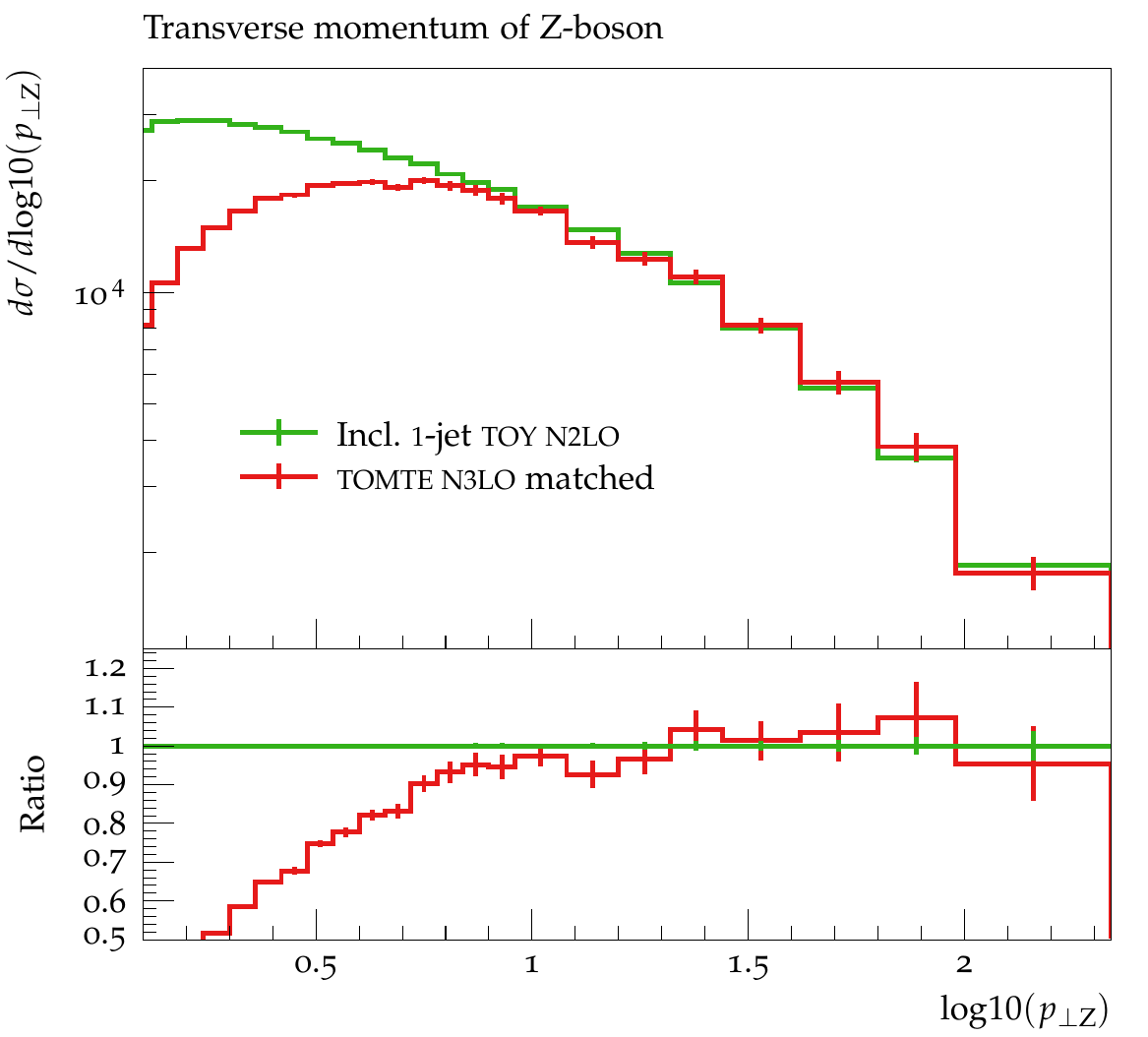}{}
  \caption{\label{fig:tomte-results-dy-1} Transverse momentum of the Drell-Yan pair, as proxy for (inclusive) observables depending on at least one jet.}
  \end{subfigure}
  \hskip 2mm
  \begin{subfigure}{0.47\textwidth}
  \includegraphics[width=0.85\textwidth]{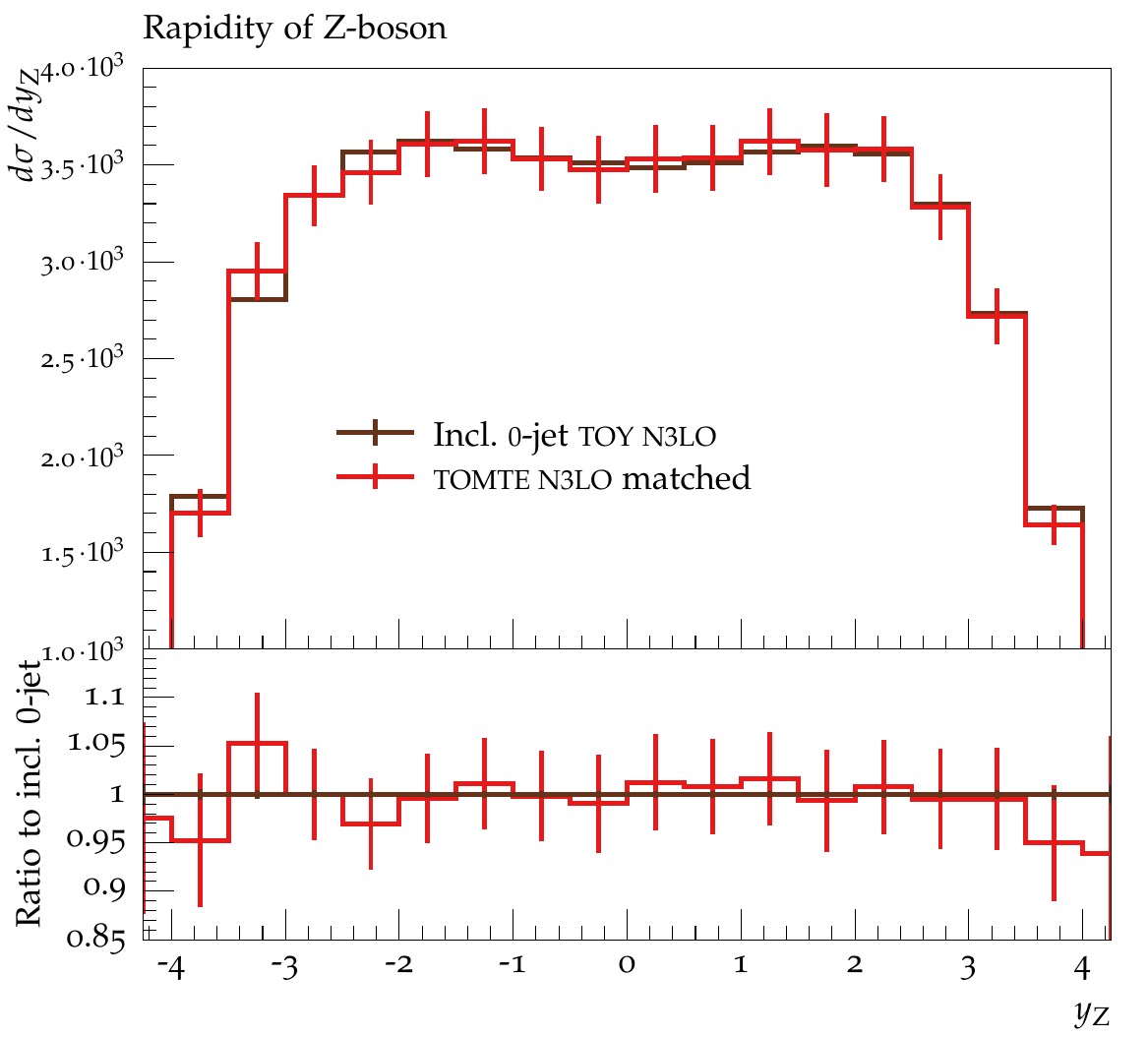}{}
  \caption{\label{fig:tomte-results-dy-0} Rapidity of the Drell-Yan pair, as proxy for observables that are fully inclusive w.r.t.\ QCD radiation.}
  \end{subfigure}
\caption{\label{fig:tomte-results-dy} 
Comparison of toy fixed-order curves and \nnnlops{} results for Drell-Yan lepton-pair production in hadron-hadron collisions. All plots have been produced with \textsc{Rivet}~\cite{Bierlich:2019rhm}. Bars denote statistical errors.
}
\end{figure}

No \nnnlo{} fixed-order \emph{event} generators are currently publicly available -- not least owing to the fact that a prescription to generate \emph{events} (i.e.\ unique phase-space points with a bounded contribution to the cross section) typically relies on an \nklopps{} method allowing a ``modified" subtraction of infrared singularities~\cite{Frixione:2002ik,*Nason:2004rx,Frixione:2007vw,Hoeche:2011fd} that allows to treat Born-like and radiative phase space regions independently. The validation of a new matching development does, however, rely on appropriate fixed-order events. As argued in~\cite{Prestel:2021vww}, this cyclic reasoning can be overcome by validating the \nnnlopps{} method with ``toy calculations". The construction of such toy calculations should not introduce a dependence on other matching schemes. The toy calculation serves to perform a closure test of (an implementation of) the \nnnlops{} method, i.e.\ to assess if the matched calculation recovers fixed-order results appropriately in the relevant phase space regions, and yields appropriately resummed results when approaching unresolved limits. Thus, the toy calculations should exhibit typical features of higher-order calculations, yet exaggerate higher-order effects to allow for conclusive validation. This note employs the strategy outlined in ~\cite{Prestel:2021vww}, with the adjustments outlined in Appendix \ref{sec:toycalc} to make the method applicable to Drell-Yan lepton pair production in proton-proton collisions. 

The \textsc{MadGraph$5$}\_a{\smaller{MC@NLO}} event generator is employed to generate leading-order events for $\mathrm{u}\bar{\mathrm{u}} \rightarrow \mathrm{e}^+\mathrm{e}^- \,+\,0,1,2,3$ gluons, which are then used to construct the toy \nnnlo{} calculation. We restrict the calculation to $m_{ \mathrm{e}^+\mathrm{e}^-}=M_\mathrm{Z}=91.2$ GeV. Proton-proton collisions are generated at $E_\mathrm{CM}=14$ TeV. The {\smaller MMHT 2014} \lo{} \pdf{} fit~\cite{Harland-Lang:2014zoa}, interfaced using {\smaller LHADPF}~\cite{Buckley:2014ana}, provides the parton distributions, as well as the $\alpha_s$ reference value, running and flavor thresholds, throughout all aspects of the calculation. Reliable and efficient \pdf{} expansions are provided by the \texttt{APFEL++} evolution library~\cite{Bertone:2013vaa,Bertone:2017gds}. The matching is performed using the \textsc{Dire} shower framework~\cite{Hoche:2015sya} interfaced with the \textsc{Pythia} $8.2$ event generator~\cite{Sjostrand:2014zea}. Hadronization and multiple scattering effects have, for the sake of a consistent closure test, been omitted in the results shown in Figure \ref{fig:tomte-results-dy}. All code to produce the inputs as well as \nnnlops{} matched results is publicly available at \url{https://gitlab.com/n3lops/tomte}.

Sample results of the closure test are shown in Figure \ref{fig:tomte-results-dy}. The ``inclusive" results indicate fixed-order predictions, determined from a separate calculation for each panel \ref{fig:tomte-results-dy-3}, \ref{fig:tomte-results-dy-2}, \ref{fig:tomte-results-dy-1} and \ref{fig:tomte-results-dy-0}. All \nnnlops{} results are obtained from a single unified event generation. 

The description of inclusive three-parton observables is illustrated by Figure~\ref{fig:tomte-results-dy-3}, showing the separation of second- and third-hardest jets in the $k_\perp$ jet clustering algorithm~\cite{Catani:1993hr}. This verifies that the \nnnlops{} prediction provides the expected Sudakov suppression when approaching the limit of the third jet becoming unresolved $d_{23}\rightarrow 0$. The \nnnlops{} prediction overshoots the reference calculation in the well-separated (\lo{}) region. This can be explained by the dynamical renormalization scale choice in the \nnnlops{} prediction, which is inherited from the parton shower prediction. The renormalization scales $t_i$ in the parton shower obey $M_\mathrm{Z}^2=t_0>t_1>t_2>\dots>t_n$, and thus lead to systematically larger $\alpha_s$ values which become ever more apparent with increasing multiplicity. It would appear natural to reaffirm this claim by (artificially) fixing the renormalization scale. That will highlight that the dynamical factorization scale used in parton showers (and hence \nnnlops{}) also has a smaller, but still non-negligible impact. Whereas it is possible to fix the renormalization scale, fixing the factorization scale leads to an inconsistent initial-state shower evolution -- unless no-emission probabilities are also omitted. Fixing all these components would lead to an uninteresting, trivial, cross-check. In conclusion, the \nnnlops{} prediction can be argued to lead to appropriate fixed-order results in the region of three well-separated jets, even though the deviation from the fixed-scale results is appreciable. The quality of the prediction should ultimately be assessed by confronting it with data.

Similar findings apply to Figure~\ref{fig:tomte-results-dy-2}, which contains the separation of the second-hardest and hardest jets. The Sudakov suppression when approaching the one-resolved-jet region is clearly visible. The (\nlo{}) fixed-order region of two well-separated jets is again populated more within \nnnlops{}, as expected from the previous argument. The effect is slightly less pronounced than for the $d_{23}$ distribution -- again as expected from the ordering of renormalization scales discussed above.

The Drell-Yan pair transverse momentum spectrum shown in Figure \ref{fig:tomte-results-dy-1} is sensitive to the presence of one or more final-state partons. The observable beautifully shows the benefits of a matched calculation: For small $p_\perp$ values, the \nnnlops{} result exhibits the desired all-order resummed regularization, the transition to high $p_\perp$ values is smooth, and at high $p_\perp$ the fixed-order (\nnlo{}) result is recovered exactly. This exact match is possible since $t_1\rightarrow M_\mathrm{Z}^2$ as $p_\perp$ becomes large, i.e.\ dynamical renormalization and factorization scales alike approach the reference value $M_\mathrm{Z}^2$. 

 Finally, the rapidity of the lepton pair (Figure \ref{fig:tomte-results-dy-0}) should not contain any higher orders introduced by (matching to) the parton shower. Indeed the \nnnlops{} prediction recovers the \nnnlo{} toy calculation exactly\footnote{The size of the statistical error bars may warrant further explanation. The method removes higher-multiplicity contamination from lower-multiplicity inclusive results (by unitarization), a higher degree of cancellation can be expected for lower multiplicities. Assuming a fixed number of pre-calculated fixed-order events for each multiplicity, this leads to a slower statistical convergence for very inclusive measurements such as the rapidity spectrum.}. Together, the comparisons in Figures \ref{fig:tomte-results-dy-3} - \ref{fig:tomte-results-dy-0} verify that the \nnnlops{} method produces \nnnlopps{}-accurate predictions also for hadron-collider processes.  

\begin{figure}[tbp]
\begin{tikzpicture}[remember picture]
\node (central) [box,inner sep=2pt, inner ysep=2pt] {
%\begin{tikzpicture}
\includegraphics[width=0.5\textwidth]{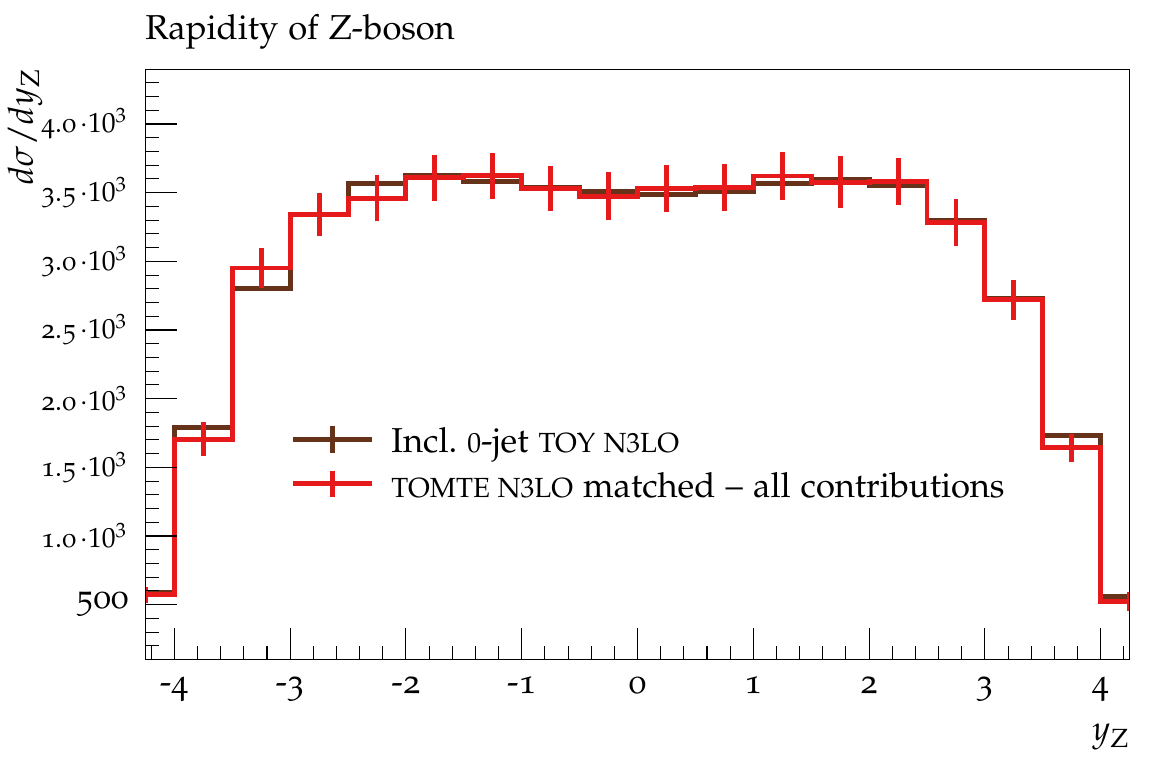}{}
%\end{tikzpicture}
};
\begin{scope}[on background layer]
\node (upleft) [box,above=of central.north west,inner sep=2pt, inner ysep=2pt, yshift=-1.25cm,xshift=-1.1cm] {
%\begin{tikzpicture}
\includegraphics[width=0.3\textwidth]{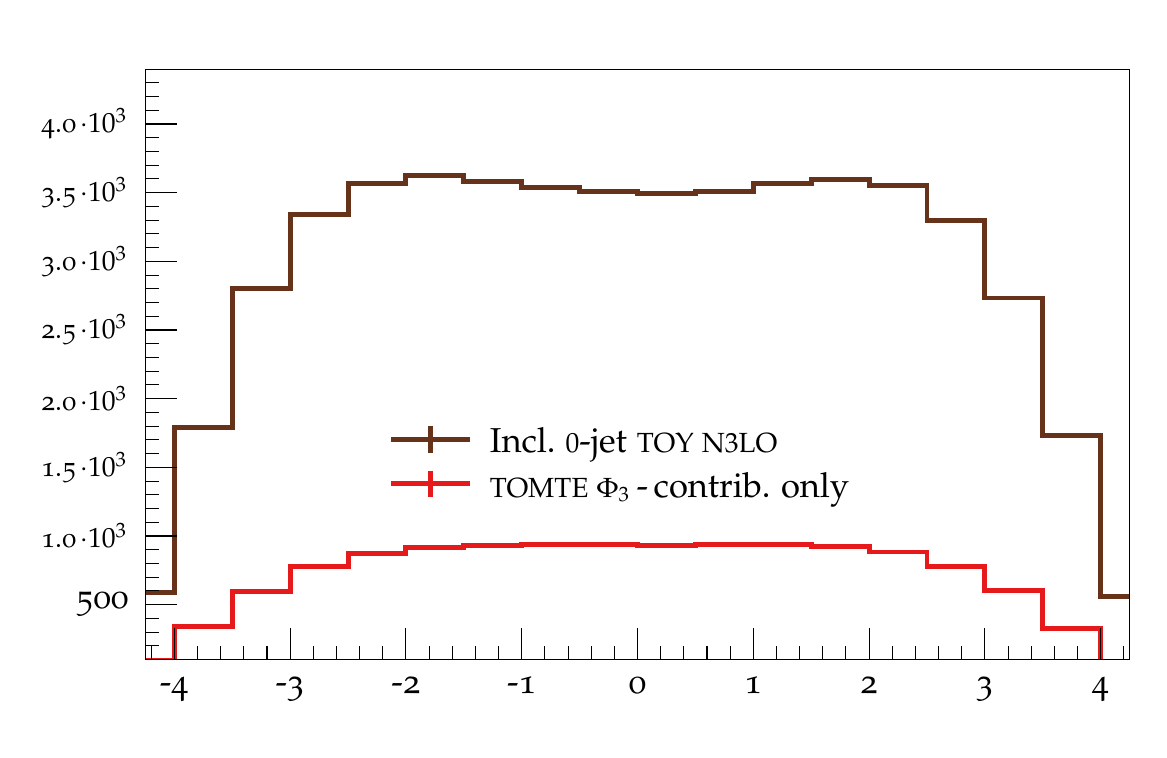}{}
%\end{tikzpicture}
};
\node (upright) [box,above=of central.north east,inner sep=2pt, inner ysep=2pt, yshift=-1.25cm,,xshift=1.1cm] {
%\begin{tikzpicture}
\includegraphics[width=0.3\textwidth]{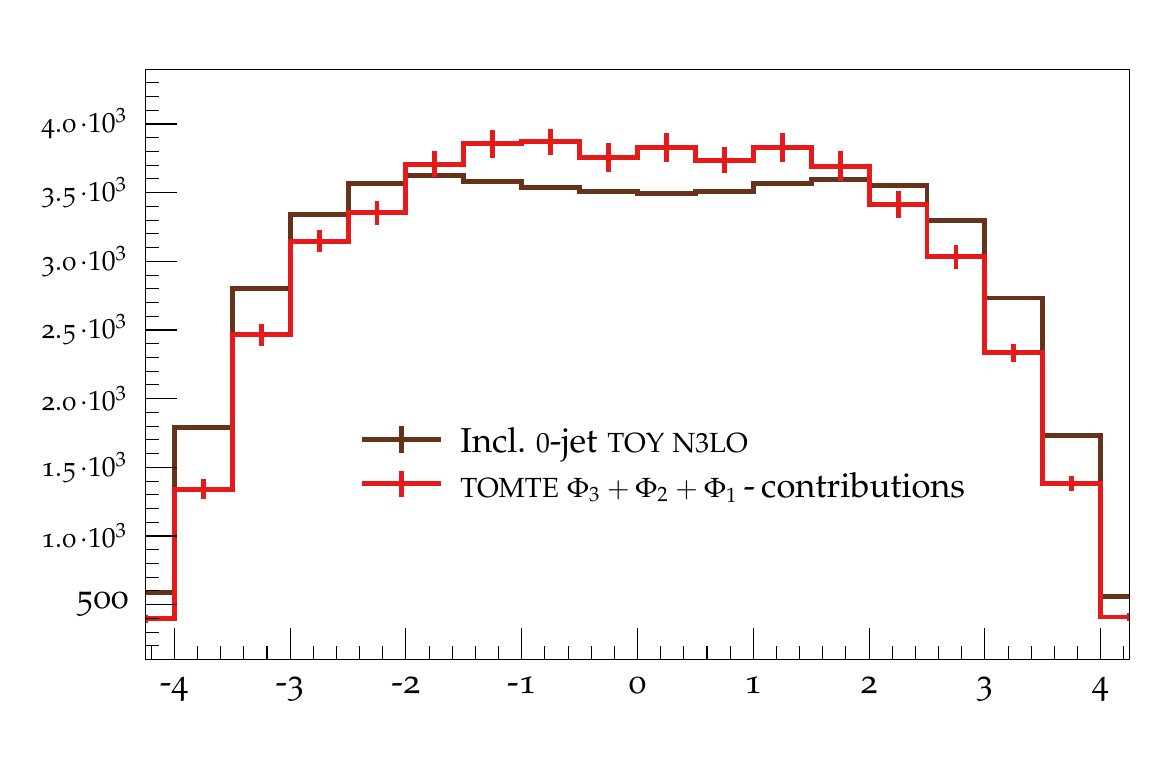}{}
%\end{tikzpicture}
};
\node (upcenter) [box,above=of central.north,inner sep=2pt, inner ysep=2pt, yshift=-1.25cm] {
%\begin{tikzpicture}
\includegraphics[width=0.3\textwidth]{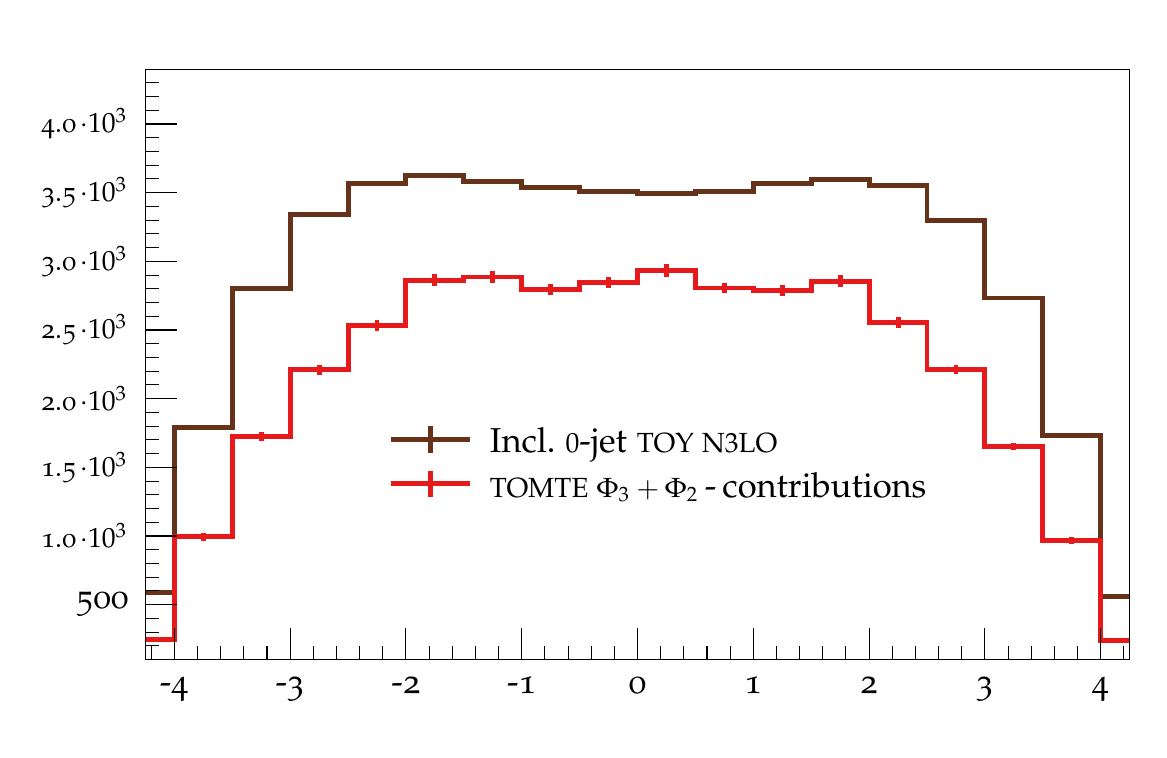}{}
%\end{tikzpicture}
};
\end{scope}
\end{tikzpicture}%
\caption{\label{fig:tomte-yz-progession} 
Comparison of various contributions to the \nnnlops{} matching formula and the baseline toy \nnnlo{} results for the Drell-Yan rapidity spectrum.
}
\end{figure}

The \nnnlops{} method combines the rates of $\Phi_{n+3},\Phi_{n+2},\Phi_{n+1}$ and $\Phi_{n+0}$ phase-space points into a consistent matched calculation. Given the many contributions to the matching formula, it may be amusing to ask: \emph{``Which contributions are really required to adequately describe an inclusive observable?"} An interesting observable in this context is the Drell-Yan rapidity spectrum, which should receive contributions from any final-state parton multiplicity. The result is shown in Figure~\ref{fig:tomte-yz-progession}. The contribution from three-partons (upper left) hardly describes the observable. Adding all two-parton terms (upper center) increases the result, though not nearly enough, whereas
further supplementing one-parton terms (upper right) overshoots the desired result at central rapidity. 
Only the combination of all contributions (containing zero up to three final-state partons partons; lower panel) leads to the correct result. Matching is required, since none of the contributions are individually close to the true distribution. 

\section{Summary and questions for the future}

Event generators form the backbone of the present and future high-energy collider program. This is especially true when relying on ``indirect searches", which rely on high-precision event generators. This note extends the \nnnlops{} method to combine \nnnlo{} \qcd{} calculations with parton showers to hadronic initial states. The reweighting in the abstract matching formula presented in \cite{Prestel:2021vww} was updated to allow for initial-state backward evolution and the dynamical factorization scale setting implied by initial-state parton showers. A numerical closure test was performed, yielding promising results: The \nnnlops{} method may be used to produce high-accuracy predictions for relevant \lhc{} processes. 
%This may be the stepping stone to a new precision frontier.

This is the first proposal for an \nnnlopps{} method capable of handling incoming hadrons. Thus, ample opportunities for future developments remain. To name a few:
\begin{itemize}
\item the treatment of configurations without parton-shower ordering requires detailed agreements between fixed-order calculations and matching frameworks; this statement applies to any matched calculation;
\item the handling of secondary or multi-parton scattering phenomena will need to be assessed carefully, both at theoretical and phenomenological level; this becomes pressing already at \nnlopps{} level;
\item processes with singularities at Born level (such as dijet production at the \lhc{}) may require further developments, such that such singularities are consistently handled in all contributions to the matching formula
\item a revised treatment of virtual fixed-order corrections that spreads their effects over higher-multiplicity phase space~\cite{Hoche:2014dla} could lead to results in closer correspondence to analytical resummation, and could be considered
%\item using antenna subtraction or similar.
\end{itemize}
We hope that these questions will inspire future work. The prototype implementation of the \nnnlops{} method used in this publication is publicly available at \url{https://gitlab.com/n3lops/tomte}.

\section*{Acknowledgement}

This note is supported by funding from the Swedish Research Council, contract numbers 2016-05996 and 2020-04303. V.~B.\ is supported by the European Union's Horizon 2020 research and innovation programme under grant agreement~824093. We thank Stefan H\"oche for positive encouragement. 

\appendix

\section{Second-order expansions of shower weights}
\label{sec:second-order-expansion}

As argued in the original \nnnlops{} publication~\cite{Prestel:2021vww}, \nnnlopps{} matching can be achieved
by considering an appropriately reweighted \nnlopps{} matched calculation, unitarization, and complementing with an \nnnlo{} exclusive (jet-vetoed) cross section. For this strategy to be successful, the main new requirement beyond aspects already present at \nnlo{} is the second-order expansion of the weight applied to tree-level contributions to the rate of one additional final state parton. The original \nnnlops{} publication was limited to uncolored initial states, in order to avoid having to consider expansions of \pdf{} factors. This is remedied below. In the presence of incoming hadrons, the expansion in $\alpha_s(\mu_\mathrm{R})$ is understood as expansion at fixed $\mu_\mathrm{F}$.

\subsection{Problem statement}
\label{sec:problemstatement}

\noindent
Overall, we need to find an appropriate subtraction $\mathcal{S}$ such that
\begin{eqnarray}
w_n \left( 1- \frac{\alpha_s}{2\pi} w_n|_1 - \left(\frac{\alpha_s}{2\pi}\right)^2 \mathcal{S} \right)~,
\end{eqnarray}
where $w_n$ is the all-order weight defined in eq.~\ref{eq:ckkwl-weight}, contains $\mathcal{O}(1)$ terms, but no $\mathcal{O}(\alpha_s)$ or $\mathcal{O}(\alpha_s^2)$ terms. All factors should be evaluated at fixed $\mu_\mathrm{F}$ and fixed $\mu_\mathrm{R}$, where applicable.

The weight $w_n$ consists of ratios of all-order factors. In general, using $x =\frac{\alpha_s}{2\pi}$, one may write
\begin{eqnarray*}
w \approx 
\frac{\left[\prod\limits_{i=1}^N a_{i0} \right]}{\left[\prod\limits_{i=1}^M b_{i0}\right]}
\frac{\left[\prod\limits_{i=1}^N (1 + x a_{i1} + x^2 a_{i2}) \right]}{\left[\prod\limits_{i=1}^M (1 + x b_{i1} + x^2 b_{i2})\right] }
\end{eqnarray*}
where $a_{i0}$ ($b_{i0}$) is the zeroth order expansion of one of the numerator (denominator factors), and 
$a_{i1}$ ($b_{i1}$) and $a_{i2}$ ($b_{i2}$) are the first and second-order expansion coefficients, divided by the zeroth-order coefficients, of the numerator (denominator) factors. It is useful to define  
\begin{eqnarray*}
c_{ij} =
\begin{cases}
a_{ij} & \mathrm{if}~ i\leq N\\
-b_{ij} & \mathrm{if}~ i> N
\end{cases}
\end{eqnarray*}
With this, 
\begin{eqnarray}
- w|_1 = -\sum\limits_{i=1}^{N+M} c_{i1}
\end{eqnarray}
and 
\begin{eqnarray*}
&&
\frac{\left[\prod\limits_{i=1}^N a_{i0} \right]}{\left[\prod\limits_{i=1}^M b_{i0}\right]}
\frac{\left[\prod\limits_{i=1}^N (1 + x a_{i1} + x^2 a_{i2}) \right]}{\left[\prod\limits_{i=1}^M (1 + x b_{i1} + x^2 b_{i2})\right] }
\left( 1 - x\sum\limits_{i=1}^{N+M} c_{i1} - x^2\mathcal{S} \right)\\
&=&
1 + x^2\left[ \sum\limits_{i=1}^{N+M} c_{i2} - \sum\limits_{i=1}^{N+M} c_{i1}  \sum\limits_{j=i}^{N+M} c_{j1} + \sum_{i=1}^M b_{i1}^2 \right] 
- x^2\mathcal{S}~.
\end{eqnarray*}
The terms $\propto b_{i1}^2$ arise due to the expansion of denominators.
Thus, the subtraction to remove second-order terms is
\begin{eqnarray}
\label{eq:2nd_order_subtraction}
\mathcal{S} &=& 
 \sum\limits_{i=1}^{N+M} c_{i2} - \sum\limits_{i=1}^{N+M} c_{i1}  \sum\limits_{j=i}^{N+M} c_{j1} + \sum_{i=1}^M b_{i1}^2
\end{eqnarray}
To construct this subtraction, the second-order expansion coefficients of all
factors in the weight need to be known.

\subsection{Running-coupling expansion}
\label{sec:expansion}

\noindent
The expansion of $\alpha_s(t)$ in terms of $\alpha_s(\mu_{\rm R})$ is straight-forward, and yields
\begin{eqnarray}
\alpha_s(t) =
  \alpha_s(\mu_{\rm R}) \left\{ 1
+  \frac{ \alpha_s(\mu_{\rm R})}{2\pi} \beta_0 \ln\left(\frac{\mu_\mathrm{R}}{t}\right)
+  \left[\frac{ \alpha_s(\mu_{\rm R})}{2\pi} \beta_0 \ln\left(\frac{\mu_\mathrm{R}}{t}\right)\right]^2
+  \left[\frac{ \alpha_s(\mu_{\rm R})}{2\pi}\right]^2 \beta_1 \ln\left(\frac{\mu_\mathrm{R}}{t}\right)\right\}
\end{eqnarray}
with
\begin{eqnarray*}
\beta_0 = \frac{11}{6}C_\mathrm{A} - \frac{2}{3}N_\mathrm{F}T_\mathrm{R} \quad,\qquad 
\beta_1 = \frac{17}{6}C_\mathrm{A}^2 - \left( \frac{5}{3} C_\mathrm{A} + C_\mathrm{F} \right)N_\mathrm{F}T_\mathrm{R} 
\end{eqnarray*}
The first-order expansion term will be needed below, as both \pdf{} evolution and no-emission probabilities rely on running-coupling evaluations.

\subsection{PDF expansion}

We now move to the second-order expansion of \pdf{} ratios. The strategy will be to extract $\mathcal{O}(\alpha_s^2(\mu_\mathrm{R}))$ terms at fixed factorization scale $\mu_\mathrm{F}$\footnote{An efficient pre-calculation of \pdf{} convolutions upon initialization currently limits the implementation to processes with ``constant natural factorization scale", and thus does not satisfactorily cover processes like e.g.\ Deep Inelastic Scattering.}. The \lo{} (\nlo{}) \dglap{} splitting kernels will be written as $\lodglapkernel{a}{b}{z}$ ($\nlodglapkernel{a}{b}{z}$).

The second-order expansion of $f_a(x,Q)$ (with $Q$ and $\mu_\mathrm{F}$ GeV$^2$-valued, and assuming the phase-space limits $C(x)$, typically $C(x)=\{z\geq x \cap z\leq1\}$) is
\begin{eqnarray}
  \label{eq:pdfevolutionexpansionLOandNLO}
f(x,Q) &=&
f (x,\mu_\mathrm{F}) \nonumber\\
%&-& \frac{\alpha_s(\mu_\mathrm{R})}{4\pi}\ln\left(\frac{\mu_\mathrm{F}}{Q}\right) P^{(0)}\otimes f(\mu_\mathrm{F})\\
&-&
\left(\frac{\alpha_s(\mu_\mathrm{R})}{4\pi}\right)^1\ln\left(\frac{\mu_\mathrm{F}}{Q}\right)
\int\displaylimits_{C(x)}  dz
\sum\limits_{b=q,g} \lodglapkernel{a}{b}{z}  \xf{b}{\frac{x}{z}}{\mu_\mathrm{F}}\\
%&-&
%\left(\frac{\alpha_s(\mu_\mathrm{R})}{4\pi}\right)^2\ln\left(\frac{\mu_\mathrm{F}}{Q}\right) P^{(1)}\otimes f(\mu_\mathrm{F})
       \label{eq:pdfevolutionexpansionNNLO}
&-&
\left(\frac{\alpha_s(\mu_\mathrm{R})}{4\pi}\right)^2\ln\left(\frac{\mu_\mathrm{F}}{Q}\right)
\int\displaylimits_{C(x)}  dz
\sum\limits_{b=q,g} \nlodglapkernel{a}{b}{z}  \xf{b}{\frac{x}{z}}{\mu_\mathrm{F}}\\
%&+&
%\left(\frac{\alpha_s(\mu_\mathrm{R})}{4\pi}\right)^2 \left[ \frac{1}{2} \ln\left(\frac{\mu_\mathrm{F}}{Q}\right)  - \ln\left(\frac{\mu_\mathrm{R}}{Q}\right)  \right] \beta_0^\mathrm{PDF} P^{(0)}\otimes f(\mu_\mathrm{F})
&+&
\left(\frac{\alpha_s(\mu_\mathrm{R})}{4\pi}\right)^2 \left[ \ln\left(\frac{\mu_\mathrm{F}}{Q}\right)  - 2\ln\left(\frac{\mu_\mathrm{R}}{Q}\right)  \right] \beta_0 
\int\displaylimits_{C(x)}  dz
\sum\limits_{b=q,g} \lodglapkernel{a}{b}{z}  \xf{b}{\frac{x}{z}}{\mu_\mathrm{F}}\nonumber\\
%&+& 
%\left(\frac{\alpha_s(\mu_\mathrm{R})}{4\pi}\right)^2 \frac{1}{2} \ln^2\left(\frac{\mu_\mathrm{F}}{Q}\right) P^{(0)}\otimes P^{(0)}\otimes f(\mu_\mathrm{F})
&+&  
\left(\frac{\alpha_s(\mu_\mathrm{R})}{4\pi}\right)^2 \frac{1}{2} \ln^2\left(\frac{\mu_\mathrm{F}}{Q}\right)
\int\displaylimits_{C(x)}  dz 
\sum\limits_{b=q,g}  \lodglapkernel{a}{b}{z} 
\int\displaylimits_{C(x/z)}  dz'
 \sum\limits_{c=q,g} \lodglapkernel{b}{c}{z'}  \xf{c}{\frac{x}{zz'}}{\mu_\mathrm{F}}\nonumber
\quad+\quad
\mathcal{O}(\alpha_s^3(\mu_\mathrm{R}))\,.\nonumber
\end{eqnarray}

\begin{table}[t!]
\renewcommand{\arraystretch}{1.5}
\begin{tabular}{ |c c c c c c|}
\hline
$a_2^{q\bar q}$~=~  $-50$, \qquad\qquad & $a_2^{q g_1}$~=~ 10, \qquad\qquad & $a_2^{q g_2}$~=~ 50, \qquad\qquad & $a_2^{\bar q g_1}$~=~ 10, \qquad\qquad & $a_2^{\bar q g_2}$~=~ 50, \qquad\qquad & $a_2^{gg}$~=~ 100 \\
$a_1^{q\bar q}$~=~ 50, \qquad\quad &$a_1^{q g}$~=~ 50, \qquad\quad &$a_1^{\bar q g}$~=~ 50, \qquad\quad &$b_1^{q\bar q}$~=~ 0, \qquad\quad &$b_1^{q g}$~=~ $-300$, \qquad\quad &$b_1^{\bar q g}$~=~ $-300$ \\
$a_0^{q}$~=~ 2, \qquad\quad &$a_0^{\bar q}$~=~ 10, \qquad\quad &$b_0^{q}$~=~ $-100$, \qquad\quad &$b_0^{\bar q}$~=~ $-100$, \qquad\quad &$c_0^{q}$~=~ 500, \qquad\quad &$c_0^{\bar q}$~=~ 500 \\
\hline
\end{tabular}
\caption{\label{tab:toycalc-coefficients} Values of the coefficients $a$, $b$ and $c$ used in eqs.\ 3.2 and 3.4 of~\cite{Prestel:2021vww} and eq.~\ref{eq:toynnnlo} of this manuscript.}
\end{table}

\subsection{No-emission probability expansion}

Next, it is necessary to expand the no-emission probabilities to second order. For splittings that do not affect initial-state particles, all details were given in~\cite{Prestel:2021vww}. 
If a radiating system $r_i$ may induce a splitting $s(r_i)$ that changes a pre-branching initial-state particle $p(s)$, thereby changing it to $p'(s)$, then the no-emission probability $\Pi_{r_i}(x; t_0, t_1)$ will contain the ratio of \pdf{}s with different momentum fraction. The expansion of such ratios is given by
\begin{eqnarray*}
 && \frac{\sum\limits_{s(r_i)} \pskernel{f}{f'(s)}{z} \xf{p'(s)}{\frac{x}{z}}{\rho}}{\xf{p(s)}{x}{\rho}}
=
\sum\limits_{s(r_i)} \pskernel{f}{f'(s)}{z} \Bigg(~~
 \frac{\xf{p'(s)}{\frac{x}{z}}{\mu_\mathrm{F}}}{\xf{p(s)}{x}{\mu_\mathrm{F}}}\\
&&
%\qquad
\quad
-\frac{\alpha_s(\mu_\mathrm{R})}{4\pi}
\ln\left(\frac{\mu_\mathrm{F}}{\rho}\right)
\frac{\xf{p'(s)}{\frac{x}{z}}{\mu_\mathrm{F}}}{\xf{p(s)}{x}{\mu_\mathrm{F}}}%\\
%&&
%\qquad\qquad\qquad~\cdot\,
\Bigg[~
\mksmall{\int\displaylimits_{C(x/z)}  dz' 
\frac{ \sum\limits_{c=q,g} \lodglapkernel{p'(s)}{c}{z'} \xf{c}{\frac{x}{zz'}}{\mu_\mathrm{F}} }
     { \xf{p'(s)}{\frac{x}{z}}{\mu_\mathrm{F}}}
 -
\int\displaylimits_{C(x)}  dz' 
\frac{  \sum\limits_{c=q,g} \lodglapkernel{p(s)}{c}{z'} \xf{c}{\frac{x}{z'}}{\mu_\mathrm{F}}}{\xf{p(s)}{x}{\mu_\mathrm{F}}}
}~
\Bigg]
~
\Bigg)
\end{eqnarray*}
With this, the second-order expansion of a no-emission probability amounts to
\begin{eqnarray}
&&
%\noem{0\pm}{x}{t_0}{t_1}
\Pi_{r_i}(x; t_0, t_1)
= 1
\nonumber\\
&&
\quad-
\int\displaylimits_{t_1}^{t_0} \frac{d\rho}{\rho} 
\sum\limits_{s(r_i)} 
%\int\displaylimits_{\Omega_{0\pm,r,s}(x,\rho)}
\int\displaylimits_{\Omega(s(r_i), \rho)}
 dz
\frac{\alpha_s(\mu_\mathrm{R})}{2\pi}
%\lopskernel{0\pm}{b(s)}{z}
\lopskernel{f}{f'(s)}{z}
\frac{\xf{p'(s)}{\frac{x}{z}}{\mu_\mathrm{F}}}{\xf{p(s)}{x}{\mu_\mathrm{F}}} 
\\
&&
\quad-
\int\displaylimits_{t_1}^{t_0} \frac{d\rho}{\rho}
\sum\limits_{s(r_i)} 
\int\displaylimits_{\Omega(s(r_i), \rho)} dz
\left(\frac{\alpha_s(\mu_\mathrm{R})}{2\pi}\right)^2
\lopskernel{f}{f'(s)}{z}
\frac{\xf{p'(s)}{\frac{x}{z}}{\mu_\mathrm{F}}}{\xf{p(s)}{x}{\mu_\mathrm{F}}} ~\Bigg[
\beta_0\ln\left(\frac{\mu_\mathrm{R}}{t}\right)\\
&&\quad\quad
-\frac{1}{2}
\ln\left(\frac{\mu_\mathrm{F}}{\rho}\right)
\Bigg\{
\mksmall{\int\displaylimits_{C(x/z)}  dz' 
\frac{ \sum\limits_{c=q,g} \lodglapkernel{p'(s)}{c}{z'} \xf{c}{\frac{x}{zz'}}{\mu_\mathrm{F}} }
     { \xf{p'(s)}{\frac{x}{z}}{\mu_\mathrm{F}}}
 -
\int\displaylimits_{C(x)}  dz' 
\frac{  \sum\limits_{c=q,g} \lodglapkernel{p(s)}{c}{z'} \xf{c}{\frac{x}{z'}}{\mu_\mathrm{F}}}{\xf{p(s)}{x}{\mu_\mathrm{F}}}
}
\Bigg\}
\Bigg]
\nonumber\\
&&
\quad-
\int\displaylimits_{t_1}^{t_0} \frac{d\rho}{\rho} 
\sum\limits_{s(r_i)} 
\int\displaylimits_{\Omega(s(r_i), \rho)} dz
\left(\frac{\alpha_s(\mu_\mathrm{R})}{2\pi}\right)^2
\frac{ 
%\nlopskernel{0\pm}{b(s)}{z}
\nlopskernel{f}{f'(s)}{z}
 \xf{p'(s)}{\frac{x}{z}}{\mu_\mathrm{F}}}{\xf{p(s)}{x}{\mu_\mathrm{F}}}
\nonumber\\&&
\quad+
\frac{1}{2}
\left(\frac{\alpha_s(\mu_\mathrm{R})}{2\pi}\right)^2
\left(
\int\displaylimits_{t_1}^{t_0} \frac{d\rho}{\rho}
\sum\limits_{s(r_i)}
\int\displaylimits_{\Omega(s(r_i), \rho)} dz
\frac{  
%\lopskernel{0\pm}{b(s)}{z} 
\lopskernel{f}{f'(s)}{z}
\xf{p'(s)}{\frac{x}{z}}{\mu_\mathrm{F}}}{\xf{p(s)}{x}{\mu_\mathrm{F}}} 
\right)^2
\quad+\quad
\mathcal{O}(\alpha_s^3(\mu_\mathrm{R}))
\nonumber
\end{eqnarray}
This completes the derivation of all ingredients required to construct the second-order subtractions $\mathcal{S}$ of eq.~\ref{eq:2nd_order_subtraction}.

\section{Constructing a toy calculation for validation}
\label{sec:toycalc}

Currently, no \nnnlo{} event generators are available. Thus, to test the feasibility and correctness of the \nnnlops{} scheme, ``toy calculations" are employed. These calculation are constructed from (minimally regularized) tree-level calculations through reweighting. Although not providing accurate results, this has the benefit that the matched calculation can be validated in detail, since the consequences of (toy) higher-order corrections are known exactly.

The method to produce a toy calculation in~\cite{Prestel:2021vww} almost directly applies to the hadron collider case. The functional form of the \nnnlo{} part of the toy calculation is modified to
\begin{eqnarray}
\label{eq:toynnnlo}
&&
\ss{0}{0+1+2+3}{\mathrm{TOY}}{\Phi_{0}}
=
\Bigg\{\i{0}\s{0}{0}{\Phi_{0}} \cdot \left[ 1
 + \frac{\alpha_s}{2\pi} \left(a_0^{q}x_{q} + a_0^{\bar q}x_{\bar q}\right)
 + \left(\frac{\alpha_s}{2\pi}\right)^2 \left( b_0^{q} \left(1-x_{q}\right)\ln{x_{q}} + b_0^{\bar q}\left(1-x_{\bar q}\right)\ln{x_{\bar q}} \right)\right.\nonumber\\
&&\left.
\qquad\qquad\qquad\qquad\qquad\qquad\qquad\qquad\qquad\quad~~\,\,
 + \left(\frac{\alpha_s}{2\pi}\right)^3 \left(c_0^{q}x_{q}\cos{2\pi x_{q}} + c_0^{\bar q}x_{\bar q}\sin{2\pi x_{\bar q}} \right)   \right]\nonumber\\
&&
\qquad\qquad\qquad\qquad\quad
 ~-~\i{1} \ss{1}{0+1+2}{\mathrm{TOY\, INC}}{\Phi_{1}} ~\Bigg\}\obs{0}
 ~+~\i{1} \ss{2}{0+1+2}{\mathrm{TOY\, INC} , Q(\Phi_{1})<Q_c }{\Phi_{1}} ~\obs{0}
~,
\end{eqnarray}
where $x_q$ ($x_{\bar q}$) is the momentum fraction of the incoming quark (anti-quark). The coefficients of the kinematic modulations (section 3 of~\cite{Prestel:2021vww}) that define all toy calculations are listed in Table~\ref{tab:toycalc-coefficients}. 

The only other additional complication arises due to \pdf{} factors. Approximations for loop integrals are constructed from higher-multiplicity matrix elements. These proxies should be evaluated using parton luminosities applicable to Born phase space points. It is assumed that the same holds for unresolved real-emission corrections. Hence, all contributions to the toy calculation for $\Phi_n$ should be evaluated with parton luminosities related to $\Phi_n$. In practise, this is achieved by applying a \pdf{} reweighting to higher-multiplicity events when constructing approximate virtual and (unresolved) real contributions.

\bibliography{ref}{}

\begin{thebibliography}{10}

\bibitem{Buckley:2011ms}
A.~Buckley {\em et~al.},
\newblock Phys. Rept. {\bf 504}, 145 (2011), 1101.2599.

\bibitem{Frixione:2002ik}
S.~Frixione and B.~R. Webber,
\newblock JHEP {\bf 06}, 029 (2002), hep-ph/0204244.

\bibitem{Nason:2004rx}
P.~Nason,
\newblock JHEP {\bf 11}, 040 (2004), hep-ph/0409146.

\bibitem{Frixione:2007vw}
S.~Frixione, P.~Nason, and C.~Oleari,
\newblock JHEP {\bf 11}, 070 (2007), 0709.2092.

\bibitem{Lavesson:2008ah}
N.~Lavesson and L.~Lonnblad,
\newblock JHEP {\bf 12}, 070 (2008), 0811.2912.

\bibitem{Hoeche:2014aia}
S.~H{\"o}che, Y.~Li, and S.~Prestel,
\newblock Phys. Rev. {\bf D91}, 074015 (2015), 1405.3607.

\bibitem{Hoche:2014dla}
S.~H\"oche, Y.~Li, and S.~Prestel,
\newblock Phys. Rev. D {\bf 90}, 054011 (2014), 1407.3773.

\bibitem{Hamilton:2013fea}
K.~Hamilton, P.~Nason, E.~Re, and G.~Zanderighi,
\newblock JHEP {\bf 10}, 222 (2013), 1309.0017.

\bibitem{Karlberg:2014qua}
A.~Karlberg, E.~Re, and G.~Zanderighi,
\newblock JHEP {\bf 09}, 134 (2014), 1407.2940.

\bibitem{Hamilton:2015nsa}
K.~Hamilton, P.~Nason, and G.~Zanderighi,
\newblock JHEP {\bf 05}, 140 (2015), 1501.04637.

\bibitem{Alioli:2015toa}
S.~Alioli, C.~W. Bauer, C.~Berggren, F.~J. Tackmann, and J.~R. Walsh,
\newblock Phys. Rev. D {\bf 92}, 094020 (2015), 1508.01475.

\bibitem{Astill:2018ivh}
W.~Astill, W.~Bizo\'n, E.~Re, and G.~Zanderighi,
\newblock JHEP {\bf 11}, 157 (2018), 1804.08141.

\bibitem{Hoche:2018gti}
S.~H\"oche, S.~Kuttimalai, and Y.~Li,
\newblock Phys. Rev. D {\bf 98}, 114013 (2018), 1809.04192.

\bibitem{Re:2018vac}
E.~Re, M.~Wiesemann, and G.~Zanderighi,
\newblock JHEP {\bf 12}, 121 (2018), 1805.09857.

\bibitem{Monni:2019whf}
P.~F. Monni, P.~Nason, E.~Re, M.~Wiesemann, and G.~Zanderighi,
\newblock JHEP {\bf 05}, 143 (2020), 1908.06987.

\bibitem{Monni:2020nks}
P.~F. Monni, E.~Re, and M.~Wiesemann,
\newblock Eur. Phys. J. C {\bf 80}, 1075 (2020), 2006.04133.

\bibitem{Lombardi:2020wju}
D.~Lombardi, M.~Wiesemann, and G.~Zanderighi,
\newblock (2020), 2010.10478.

\bibitem{Hu:2021rkt}
Y.~Hu, C.~Sun, X.-M. Shen, and J.~Gao,
\newblock (2021), 2101.08916.

\bibitem{Alioli:2020qrd}
S.~Alioli {\em et~al.},
\newblock (2020), 2010.10498.

\bibitem{Mazzitelli:2020jio}
J.~Mazzitelli {\em et~al.},
\newblock (2020), 2012.14267.

\bibitem{Anastasiou:2015vya}
C.~Anastasiou, C.~Duhr, F.~Dulat, F.~Herzog, and B.~Mistlberger,
\newblock Phys. Rev. Lett. {\bf 114}, 212001 (2015), 1503.06056.

\bibitem{Duhr:2019kwi}
C.~Duhr, F.~Dulat, and B.~Mistlberger,
\newblock Phys. Rev. Lett. {\bf 125}, 051804 (2020), 1904.09990.

\bibitem{Duhr:2020seh}
C.~Duhr, F.~Dulat, and B.~Mistlberger,
\newblock Phys. Rev. Lett. {\bf 125}, 172001 (2020), 2001.07717.

\bibitem{Chen:2019lzz}
L.-B. Chen, H.~T. Li, H.-S. Shao, and J.~Wang,
\newblock Phys. Lett. B {\bf 803}, 135292 (2020), 1909.06808.

\bibitem{Duhr:2020sdp}
C.~Duhr, F.~Dulat, and B.~Mistlberger,
\newblock JHEP {\bf 11}, 143 (2020), 2007.13313.

\bibitem{Dulat:2017prg}
F.~Dulat, B.~Mistlberger, and A.~Pelloni,
\newblock JHEP {\bf 01}, 145 (2018), 1710.03016.

\bibitem{Currie:2018fgr}
J.~Currie {\em et~al.},
\newblock JHEP {\bf 05}, 209 (2018), 1803.09973.

\bibitem{Dreyer:2018qbw}
F.~A. Dreyer and A.~Karlberg,
\newblock Phys. Rev. D {\bf 98}, 114016 (2018), 1811.07906.

\bibitem{Cieri:2018oms}
L.~Cieri, X.~Chen, T.~Gehrmann, E.~W.~N. Glover, and A.~Huss,
\newblock JHEP {\bf 02}, 096 (2019), 1807.11501.

\bibitem{Mondini:2019gid}
R.~Mondini, M.~Schiavi, and C.~Williams,
\newblock JHEP {\bf 06}, 079 (2019), 1904.08960.

\bibitem{Chen:2021isd}
X.~Chen {\em et~al.},
\newblock (2021), 2102.07607.

\bibitem{Billis:2021ecs}
G.~Billis, B.~Dehnadi, M.~A. Ebert, J.~K.~L. Michel, and F.~J. Tackmann,
\newblock (2021), 2102.08039.

\bibitem{Camarda:2021ict}
S.~Camarda, L.~Cieri, and G.~Ferrera,
\newblock (2021), 2103.04974.

\bibitem{Re:2021con}
E.~Re, L.~Rottoli, and P.~Torrielli,
\newblock (2021), 2104.07509.

\bibitem{Banfi:2015pju}
A.~Banfi {\em et~al.},
\newblock JHEP {\bf 04}, 049 (2016), 1511.02886.

\bibitem{Prestel:2021vww}
S.~Prestel,
\newblock JHEP {\bf 11}, 041 (2021), 2106.03206.

\bibitem{Amoroso:2020lgh}
S.~Amoroso {\em et~al.},
\newblock {Les Houches 2019: Physics at TeV Colliders: Standard Model Working
  Group Report},
\newblock in {\em {11th Les Houches Workshop on Physics at TeV Colliders}:
  {PhysTeV Les Houches}}, 2020, 2003.01700.

\bibitem{Gottschalk:1986bk}
T.~D. Gottschalk,
\newblock Nucl. Phys. B {\bf 277}, 700 (1986).

\bibitem{Nagy:2020gjv}
Z.~Nagy and D.~E. Soper,
\newblock Phys. Rev. D {\bf 102}, 014025 (2020), 2002.04125.

\bibitem{Rubin:2010xp}
M.~Rubin, G.~P. Salam, and S.~Sapeta,
\newblock JHEP {\bf 09}, 084 (2010), 1006.2144.

\bibitem{Lonnblad:2012ng}
L.~L{\"o}nnblad and S.~Prestel,
\newblock JHEP {\bf 02}, 094 (2013), 1211.4827.

\bibitem{Lonnblad:2001iq}
L.~L{\"o}nnblad,
\newblock JHEP {\bf 05}, 046 (2002), hep-ph/0112284.

\bibitem{Lonnblad:2011xx}
L.~L{\"o}nnblad and S.~Prestel,
\newblock JHEP {\bf 03}, 019 (2012), 1109.4829.

\bibitem{Sjostrand:2004ef}
T.~Sj{\"o}strand and P.~Z. Skands,
\newblock Eur. Phys. J. {\bf C39}, 129 (2005), hep-ph/0408302.

\bibitem{Schumann:2007mg}
S.~Schumann and F.~Krauss,
\newblock JHEP {\bf 03}, 038 (2008), 0709.1027.

\bibitem{Platzer:2009jq}
S.~Platzer and S.~Gieseke,
\newblock JHEP {\bf 01}, 024 (2011), 0909.5593.

\bibitem{Hoche:2015sya}
S.~H{\"o}che and S.~Prestel,
\newblock Eur. Phys. J. {\bf C75}, 461 (2015), 1506.05057.

\bibitem{Sjostrand:1985xi}
T.~Sj{\"o}strand,
\newblock Phys. Lett. {\bf 157B}, 321 (1985).

\bibitem{Lonnblad:2012ix}
L.~L{\"o}nnblad and S.~Prestel,
\newblock JHEP {\bf 03}, 166 (2013), 1211.7278.

\bibitem{Sjostrand:1987su}
T.~Sjostrand and M.~van Zijl,
\newblock Phys. Rev. D {\bf 36}, 2019 (1987).

\bibitem{Bierlich:2019rhm}
C.~Bierlich {\em et~al.},
\newblock SciPost Phys. {\bf 8}, 026 (2020), 1912.05451.

\bibitem{Hoeche:2011fd}
S.~Hoeche, F.~Krauss, M.~Schonherr, and F.~Siegert,
\newblock JHEP {\bf 09}, 049 (2012), 1111.1220.

\bibitem{Harland-Lang:2014zoa}
L.~A. Harland-Lang, A.~D. Martin, P.~Motylinski, and R.~S. Thorne,
\newblock Eur. Phys. J. C {\bf 75}, 204 (2015), 1412.3989.

\bibitem{Buckley:2014ana}
A.~Buckley {\em et~al.},
\newblock Eur. Phys. J. {\bf C75}, 132 (2015), 1412.7420.

\bibitem{Bertone:2013vaa}
V.~Bertone, S.~Carrazza, and J.~Rojo,
\newblock Comput. Phys. Commun. {\bf 185}, 1647 (2014), 1310.1394.

\bibitem{Bertone:2017gds}
V.~Bertone,
\newblock PoS {\bf DIS2017}, 201 (2018), 1708.00911.

\bibitem{Sjostrand:2014zea}
T.~Sj{\"o}strand {\em et~al.},
\newblock Comput. Phys. Commun. {\bf 191}, 159 (2015), 1410.3012.

\bibitem{Catani:1993hr}
S.~Catani, Y.~L. Dokshitzer, M.~H. Seymour, and B.~R. Webber,
\newblock Nucl. Phys. B {\bf 406}, 187 (1993).

\end{thebibliography}
\bibliographystyle{h-physrev}
\end{document}